\documentclass[twocolumn,showpacs,preprintnumbers,amsmath,amssymb]{revtex4}
\usepackage{graphicx,epsfig}
\usepackage{dcolumn}
\usepackage{bm}
\newcommand{\ba}{\begin{eqnarray}}
\newcommand{\ea}{\end{eqnarray}}
\newcommand{\bse}{\begin{subequations}}
\newcommand{\ese}{\end{subequations}}

\newcommand{\DD}{{\cal {D}}}

\newcommand{\bbq}{\begin{quote}}
\newcommand{\eeq}{\end{quote}}
\newcommand{\tbb}{t_{\textrm{\tiny{bb}}}}
\newcommand{\tcoll}{t_{\textrm{\tiny{coll}}}}
\newcommand{\tmax}{t_{\textrm{\tiny{max}}}}
\newcommand{\Lmax}{L_{\textrm{\tiny{max}}}}
\newcommand{\RR}{{}^3{\cal{R}}}
\newcommand{\RRi}{{}^3{\cal{R}}_i}
\newcommand{\T}{{}^3{\cal{T}}}

\newcommand{\EE}{{\cal{E}}}
\newcommand{\FF}{{\cal{F}}}

\newcommand{\VV}{{\cal{V}}}
\newcommand{\HH}{{\cal{H}}}

\newcommand{\hOm}{\hat\Omega}
\newcommand{\hOmi}{\hat\Omega_i}

\newcommand{\Dih}{\delta_i^{(\HH)}}
\newcommand{\Dim}{\delta_i^{(m)}}
\newcommand{\Dik}{\delta_i^{(k)}}

\newcommand{\Da}{\delta^{(A)}}

\newcommand{\Dh}{\delta^{(\HH)}}

\newcommand{\Dm}{\delta^{(m)}}
\newcommand{\dDm}{\dot\delta^{(m)}}
\newcommand{\Dk}{\delta^{(k)}}

\newcommand{\dd}{{\rm{d}}}
\newcommand{\rtv}{r_{\rm{tv}}}

\begin{document}

\title{A new approach for doing theoretical and numeric work with Lema\^{\i}tre--Tolman--Bondi dust models. } 
\author{ Roberto A. Sussman$^\ddagger$}
\affiliation{
$^\ddagger$Instituto de Ciencias Nucleares, Universidad Nacional Aut\'onoma de M\'exico (ICN-UNAM),
A. P. 70--543, 04510 M\'exico D. F., M\'exico. }
%
%
\date{\today}
\begin{abstract} We introduce quasi--local integral scalar variables for the study of spherically symmetric Lema\^{i}tre--Tolman--Bondi (LTB) dust models.  Besides providing a covariant, and theoretically appealing,  interpretation for the parameters of these models, these variables allow us to study their dynamics (in their full generality) by means of fluid flow evolution equations that can be handled with simple numerical techniques and has a significant potential for astrophysical and cosmological applications. These evolution equations can also be understood in the framework of a gauge invariant and covariant formalism of spherical non--linear perturbations on a FLRW background. The covariant time splitting associated the new variables leads, in a natural way, to rephrase the known analytic solutions within an initial value framework in which covariant scalars are given by simple scaling laws. By using this re--parametrization of the analytic solutions, we re--examine and provide an alternative outlook to various theoretical issues already treated in the literature: regularity conditions, an Omega parameter, as well as the fitting of a given LTB model to radial profiles of density or velocity at different cosmic times. Other theoretical issues and numeric applications will be examined in separate articles.                             
\end{abstract}

\maketitle
\section{Introduction.}

Spherically symmetric LTB dust models are among the best known exact solutions of Einstein's equations \cite{LTB}. Since they provide access to non--linear effects associated with inhomogeneous sources by means of analytic and/or tractable numeric solutions, they have received widespread attention in the literature (see \cite{kras,kras2} for comprehensive reviews). These exact solutions are used mostly to study cosmological inhomogeneities, as for example the series of articles in the references \cite{KH1,KH2,KH3,KH4}, but have also been employed to examine a variety of issues, for example as useful toy models to describe gravitational collapse under a classical \cite{lemos,joshi} and quantum \cite{quantum} approach. LTB models have also been widely used \cite{LTB1,LTB2,LTBkolb,LTBchin,LTBfin,num3,num2,num1} in the context of the ongoing widespread theoretical and empiric effort to explore the possibility that cosmic observations could be explained without resorting to dark energy, but by taking into consideration the fully non--linear effects of inhomogeneities in the context of General Relativity (see \cite{celerier} for a review). These models are also helpful as theoretical tools to probe various averaging techniques applied to cosmological inhomogeneities \cite{LTBchin,LTBfin,LTBave1,LTBave2,LTBave3}, in particular, the scalar averaging formalism developed by Buchert and coworkers \cite{ave_review}. One finds in the extensive literature on these models a preferred set of free functions and analytic solutions (implicit and parametric) that has become a sort of standard \cite{kras,kras2,ltbstuff} (even in numeric work on these models \cite{LTBchin,LTBfin,num3,num2,num1}). 

Given the success of these models and the fact that their standard parametrization is adequate and works in practice, then it is legitimate to ask for the justification of an article proposing and discussing alternative variables. The answer to this question is simple: alternative variables may provide or motivate either new theoretical and empiric developments or allow us to grasp known results under a different perspective, thus illuminating possible unsuspected connections with other models or theoretical issues. In this context, we propose an alternative description of LTB dust models that is based on a set of covariant scalar variables defined by applying to arbitrary scalars the same type of integral construct that yields the Misner--Sharp quasi--local mass function \cite{MSQLM,hayward1,hayward2}. Hence, we find it natural to call these variables quasi--local scalars.  

The quasi--local scalars mentioned above have already been used for a dynamical system study of dust LTB models \cite{suss08} and in the application of Buchert's averaging formalism to LTB models \cite{LTBave2,LTBave3}, as well as  in a numeric study of inhomogeneous dark energy sources associated with an LTB metric, but with an energy--momentum tensor that is more general than dust \cite{sussQL,suss2009}. The pure dust case merits the separate study that we provide here, not only because dust is an adequate source to model cold dark matter in cosmological scales (and a much less artificial source for an LTB metric than the fluid tensor in \cite{sussQL,suss2009}), but because dust sources allow for a qualitative analytic framework that complements numeric work and yields very useful theoretical and practical information. 

The potential of the quasi--local variables for analytic and qualitative work on LTB models follows readily from parametrizing their known analytic expressions in terms of these  variables and their fluctuations. The new variables lead in a natural way to an initial value parametrization of LTB models, which in turn leads to the introduction of a FLRW--like metric and scaling laws for density, Hubble expansion, spatial curvature, shear and all other covariant  quantities of physical or geometric relevance. Various elements of this parametrization have been already considered in the literature \cite{LTBkolb,LTBchin,LTBfin,num2,num3}, but only as a useful coordinate ansatzes that are justified because the involved expressions resemble FLRW parameters. In this article we provide an appealing theoretical and covariant context for these expressions that could allow for new empiric or numeric results and/or a better understanding of existing work.

The quasi--local scalar representation applied the LTB dust case leads to a complete description of the dynamics in the context of ``fluid flow'' (or ``1+3'') evolution equations that are fully general, but can be handled by simple numerical methods, and thus have a significant potential for empiric modeling of cosmological dust sources. This scalar representation also leads to a description of LTB dust models in the framework of spherical, non--linear, gauge invariant and covariant (GIC) perturbations  over a FLRW background (in the context of the fluid flow perturbation formalism derived by Ellis and Bruni \cite{ellisbruni89,BDE,1plus3}). The resulting perturbation framework is also consistent with the traditional perturbation approach \cite{bardeen}, as the evolution equations of quasi--local scalars and their fluctuations generalize the evolution equations of linear perturbations of spherical dust sources in the comoving gauge \cite{padma}.      

The plan of the article is described below. We summarize in section II the metric, field equations, analytic solutions and basic regularity of LTB models in the conventional variables. In section III we introduce the representation of local covariant scalars and their evolution equations in the context of the covariant time slicing associated with a fluid flow  approach to the dynamics of the models.  The quasi--local scalars and their fluctuations are introduced and discussed in section IV, while section V deals with some theoretical and practical implications of these scalars. In particular, the quasi--local scalars lead naturally to an initial value approach in which initial value functions are given by the new variables evaluated at a fiducial (initial) hypersurface defined by the 1+3 time slicing. This approach leads to a FLRW--like form for the LTB metric, which has been suggested previously (see \cite{LTBkolb,LTBchin}) but as an ansatz without a theoretical context. The initial value approach also leads to simple scaling laws for all scalars and an intuitive gauge for the radial coordinate. Hence, in section VI we rewrite the analytic solutions of the models in terms of the above mentioned initial value parametrization, which leads in section VII to an elegant characterization of the coordinate locus of the curvature singularities that is simpler and more intuitive than that using conventional variables. In section VIII we rewrite the Hellaby--Lake conditions \cite{ltbstuff,ltbstuff1} to avoid shell--crossings in terms of the initial value functions, while in section IX we rewrite the analytic solutions in the same initial value approach, but using a suitably defined Omega parameter and the quasi--local expansion factor as initial value functions (which can be used also to parametrize the Hellaby--Lake conditions). 

In section X we address (as an application) an interesting theoretical issue raised by Hellaby and Krasinski (see \cite {KH1,KH4,num1}), namely: the possibility of mapping by an LTB model of a given density or velocity radial profile at one cosmic time to another density or velocity profile at a latter time. We show how the initial value parametrization of the analytic solutions greatly simplifies the understanding and handling of this problem. In section XI we derive three equivalent systems of fluid flow evolution equations for the quasi--local scalars and their fluctuations. Each of these systems provides a full description of the dynamics of the models, and all are effectively equivalent to systems of ordinary differential equations subjected to constraints, hence they are technically more accessible than (and as fully general as) the system that arises for the local scalars in the 1+3 fluid flow framework. We explain how initial conditions can be set up and argue that these equations have a significant potential for numeric computation of quantities of observational interest (null geodesics, red shift factor, area and luminosity distances, etc). In section XII we show how the quasi--local scalars, their fluctuations and their evolution equations can be understood in the framework of a gauge invariant and covariant formalism of spherical non--linear perturbations. 

In section XIII we derive a fourth system of fluid flow evolution equations, but now for the Omega and Hubble parameters introduced in section IX. These evolution equations lead to a dynamical system (like the one in \cite{suss08}). We comment on the relation between the perturbation formalism of section XII and these parameters (and others proposed in the literature \cite{num2,num3}), thus providing a theoretical context for quantities given as coordinate ansatzes on the basis of their resemblance to FLRW Omega and Hubble parameters.  
We discuss in section XIV various aspects of special LTB configurations, such as: closed models, models with simultaneous big--bang or collapsing singularities, simultaneous maximal expansion and a mixed elliptic/hyperbolic configuration. Our conclusions and suggestions for future work are summarized in section XV. In Appendix A we discuss various regularity issues of the quasi--local scalars and their fluctuations, while Appendix B discusses the relation between quasi--local spatial curvature (which determines the kinematic class: parabolic, hyperbolic, elliptic) and local spatial curvature.

\section{LTB dust models in the conventional variables.}

Lemaitre--Tolman--Bondi (LTB) dust models~\cite{LTB,kras} are the spherically symmetric solutions of Einstein's equations characterized by the LTB line element and the energy--momentum tensor
\ba ds^2=-c^2dt^2+\frac{R'{}^2}{1+E}\,dr^2+R^2(d\theta^2+ 
\sin^2\theta d\varphi^2),\label{LTB1}\\
T^{ab} = \rho\,c^2\,u^a\,u^b,\qquad u^a=\delta^a_0,\label{Tab}\ea
where $R=R(t,r)$,\,\ $R'=\partial R/\partial r $,  $E=E(r)$ and $\rho=\rho(t,r)$ is the rest--mass density. The field equations reduce to
\ba \dot R^2 &=& \frac{2M}{R} +E,\label{fieldeq1}\\
 2M' &=& \kappa \rho \,R^2 \,R',\label{fieldeq2}\ea
where $\kappa=8\pi G/c^2$,\, $M=M(r)$ and $\dot R=u^a\nabla_a R=\partial R/\partial (ct)$.  

\subsection{Kinematic equivalence class and analytic solutions.} 

It is common usage in the literature  (see \cite{kras,kras2,ltbstuff}) to classify the solutions of (\ref{fieldeq1}) in ``kinematic equivalence classes'' given by the sign of $E$, which determines the existence of a zero of $\dot R^2$. Since $E=E(r)$, the sign of this function can be, either the same in the full range of $r$, in which case we have LTB models of a given kinematic class, or it can change sign in specific ranges of $r$, defining LTB models with regions of different kinematic class (see \cite{ltbstuff}).  The solutions of the Friedman--like field equation (\ref{fieldeq1}) for each kinematic class take the following well known parametric form:\\

\noindent
{\underline{Parabolic models or regions:}} \,\, $E=0$.
\begin{equation}c(t-\tbb) = \frac{2}{3}\,\eta^3,\qquad R=(2M)^{1/3}\,\eta^2,\label{par1}\end{equation}
\noindent
{\underline{Hyperbolic models or regions:}} \,\, $E\geq 0$.
\begin{equation} 
R =\frac{M}{E}\,\left(\cosh\,\eta-1\right),\qquad 
c(t-\tbb)=\frac{M}{E^{3/2}}\,\left(\sinh\,\eta-\eta\right),\label{hypt1}
\end{equation}
\noindent
{\underline{Elliptic models or regions:}} \,\, $E\leq 0$.
\begin{equation}
R =\frac{M}{|E|}\,\left(1-\cos\,\eta\right),\qquad
c(t-\tbb)=\frac{M}{|E|^{3/2}}\,\left(\eta-\sin\,\eta\right),\label{ellt1}          
\end{equation}
where $\tbb=\tbb(r)$ is called ``big bang time'', as it marks the coordinate locus of the central expanding curvature singularity: $R(t,r)=0$ for $r\geq 0$. It emerges as an ``integration constant''  in the integration of (\ref{fieldeq1}). Notice that the locus of central curvature singularity is distinct from that of the center of symmetry $R(t,0)=0$ (see Appendix \ref{symmetry_centers}). Besides the kinematic class, LTB models that admit (at least) one symmetry center can be classified as ``open'' or ``closed'', respectively corresponding to the hypersurfaces of constant $t$ being topologically equivalent to $\mathbb{R}^3$ or $\mathbb{S}^3$ (see Appendix \ref{topology}).    

In order to deal with a given specific model by means of (\ref{par1})--(\ref{ellt1}), we need to prescribe the three ``conventional'' free functions 
\begin{equation} M(r),\qquad E(r),\qquad\tbb(r).
\label{freefuncs}\end{equation}
However, the metric (\ref{LTB1}) is invariant under rescalings of the radial coordinate $r=r(\bar r)$, hence it is always possible to eliminate one of the free functions in (\ref{freefuncs}) by a suitable choice of the radial coordinate, thus leaving only two basic irreducible free functions. 

Given a choice of free functions (\ref{freefuncs}), the solutions (\ref{par1})--(\ref{ellt1}) can be used to find the remaining relevant quantities that may be required for a specific problem, a procedure that has been used abundantly in the literature \cite{kras,kras2}. 

\subsection{Regular LTB models}

We will assume henceforth the existence of (at least) one symmetry center marked by $r=0$ (see Appendix \ref{symmetry_centers}), and will denote by ``regular LTB model'' any LTB configuration for which the condition   
\begin{equation}\hbox{sign}(R')=\hbox{sign}(M')=\hbox{sign}(\sqrt{1+E}), \label{noshx} \end{equation}
holds for all $t$ and all $r$ not marking a symmetry center (since $M'(0)=0,\,E(0)=0$ and $R'(t,0)=1$ \cite{ltbstuff,suss02}). Conditions (\ref{noshx}) imply that if $R',\,M',\,\sqrt{1+E}$ have a zero under regular conditions, it must be a common zero of the same order at some worldline $r=\rtv>0$. Violation of (\ref{noshx}), {\it{i.e.}} $R'=0$ for $M'\ne 0,\,\sqrt{1+E}\ne 0$, results in a  shell crossing singularity for which $\rho\to\infty$ occurs with $R>0$. Condition (\ref{noshx}) together with (\ref{fieldeq2}) also implies that $\rho\geq 0$ holds everywhere and that it is bounded everywhere except at the coordinate locus of a central singularity ({\it{i.e.}} points for which $R(t,r)=0$ holds that are not the worldline of a symmetry center). The restrictions on the free parameters (\ref{freefuncs}) that guarantee the fulfillment of (\ref{noshx}) are the well known Hellaby--Lake (necessary and sufficient) conditions \cite{ltbstuff,suss02,ltbstuff1} given by\\ 


\noindent {\underline{Parabolic and hyperbolic models or regions:}}
\begin{equation} R'>0\quad \Leftrightarrow\quad \left\{ M'\geq 0,\quad E'\geq 0,\quad \tbb'\leq 0\right\},\label{noshx1}\end{equation}

\noindent{\underline{Elliptic models or regions:}}
\ba  \pm R' > 0\quad &\Leftrightarrow& \quad \pm  M'\geq 0,\quad \pm \tbb'\leq 0,\nonumber\\
&{}& \quad\pm\left[\frac{M'}{M}-\frac{3}{2}\frac{E'}{E}+\frac{c\tbb'\,|E|^{3/2}}{2\pi\,M}\right]\geq 0,\nonumber\\\label{noshx2}
\ea
where only expanding configurations are considered in (\ref{noshx1}) and the $\pm$ sign in (\ref{noshx2}) accounts for the fact that $R'<0$ occurs in elliptic models whose that admit a second symmetry center (see Appendix \ref{topology}). The equal sign holds only at symmetry centers and at values of $r$ where $R'=0$.
 

\section{Covariant objects and their ``1+3'' evolution equations.}

The normal geodesic 4--velocity in (\ref{LTB1})--(\ref{Tab}) defines a natural time slicing in which the space slices are the hypersurfaces $\T[t]$, orthogonal to $u^a$ and marked by arbitrary constant values of $t$\footnote{We use the notation $[t]$ to emphasize that $t$ is a fixed arbitrary parameter.}. The metric of the $\T[t]$ is simply $h_{ab}=g_{ab}+u_a u_b=g_{ij}\delta^i_a\delta^j_b$ (or (\ref{LTB1}) with $\dd t =0$). The proper volume element and 3--dimensional Ricci scalar associated with these hypersurfaces are
\ba 
\dd \VV_p &=& \sqrt{{\rm{det}}(h_{ab})}\,\dd^3x = \frac{\,R^2 R'\,\sin^2\theta\,\dd r\,\dd\theta\,\dd\varphi}{\sqrt{1+E}},\label{dV}\\
\RR &=& -\frac{2(E\,R)'}{R^2R'} =-\frac{2E'}{R'R}-\frac{2E}{R^2},\label{3Ricci1}\ea
Given this time splitting, the covariant objects in dust LTM models are $\rho$ and $\RR$ in (\ref{fieldeq2}) and (\ref{3Ricci1}), together with the shear and electric Weyl tensors, $\sigma_{ab}=\tilde\nabla_{(a}u_{b)}-(\Theta/3)h_{ab}$ and $E_{ab}u_cu_d C^{abcd}$, where  $\tilde\nabla_a = h_a^b\nabla_b$,\, and $C^{abcd}$ is the Weyl tensor, and the expansion scalar, $\Theta$, is given by
\begin{equation} 
\Theta = \tilde\nabla_au^a=\frac{2\dot R}{R}+\frac{\dot R'}{R'}.\label{Theta1}\end{equation}
Since LTB models (as all spherically symmetric spacetimes) are locally rotationally symmetric (LRS), all their covariant objects are either scalars or entirely characterized by covariant scalars~\cite{LRS}. Therefore, the spatial trace--less shear and electric Weyl tensors can be given in terms of single scalar functions, $\Sigma,\,\EE$, in a covariant manner:
\begin{equation}  
\sigma^{ab} = \Sigma\,\Xi^{ab}\quad \Rightarrow\quad \Sigma =\frac{1}{6}\,\sigma_{ab}\,\Xi^{ab}= -\frac{1}{3}\left(\frac{\dot R'}{R'}-\frac{\dot R}{R}\right),
\label{Sigma1}\end{equation}
\begin{equation}  
E^{ab} = \EE\,\Xi^{ab}\quad \Rightarrow\quad \EE =\frac{1}{6}\,E_{ab}\,\Xi^{ab}= -\frac{\kappa}{6}\rho+\frac{M}{R^3},
\label{EE1}\end{equation}
where $\Xi^{ab}\equiv h^{ab}-3\chi^{a}\chi^b$ and $\chi^a=\sqrt{h^{rr}}\,\delta^a_r$ is the unit vector orthogonal to $u^a$ and to the 2--spheres orbits of SO(3). 

A description of the dynamics of LTB models, which is an equivalent alternative to the field equations and their analytic solutions (\ref{par1})--(\ref{ellt1}), follows by considering the following set of covariant scalars    
\begin{equation} A=\left\{\rho,\,\Theta,\,\Sigma,\,\EE,\,\RR\right\},\label{locscals}\end{equation}
which completely characterize these models. Hence, their dynamics is completely determined by suitable evolution equations and constraints for these scalars. Following the ``fluid flow'' or covariant ``1+3'' framework of Ehlers, Ellis, Bruni, Dunsbury and van Ellst~\cite{ellisbruni89,BDE,1plus3,LRS}, and considering (\ref{Sigma1}) and (\ref{EE1}), the 1+3 evolution equations for the scalars (\ref{locscals}) are
\bse\label{ev_13}\ba
\dot\Theta &=&-\frac{\Theta^2}{3}
-\frac{\kappa}{2}\,\rho-6\Sigma^2,\label{ev_theta_13}\\
\dot \rho &=& -\rho\,\Theta,\label{ev_rho_13}\\
\dot\Sigma &=& -\frac{2\Theta}{3}\,\Sigma+\Sigma^2-\EE,
\label{ev_Sigma_13}\\
 \dot\EE &=& -\frac{\kappa}{2}\,\rho\Sigma
-3\,\EE \left(\frac{\Theta}{3}+\Sigma\right),\label{ev_EE_13}\ea\ese
while the spacelike and Hamiltonian constraints are
\begin{equation} \left(\Sigma+\frac{\Theta}{3}\right)'+3\,\Sigma\,\frac{R'}{R}=0,\qquad \frac{\kappa}{6}\,\rho'
+\EE\,'+3\,\EE\,\frac{R'}{R}=0,\label{c_13} \end{equation}
\begin{equation}\left(\frac{\Theta}{3}\right)^2 = \frac{\kappa}{3}\, \rho
-\frac{\RR}{6}+\Sigma^2,\label{cHam_13}\end{equation}
Solving this system of partial differential equations determines the metric functions and remaining quantities though the definitions (\ref{fieldeq1}), (\ref{fieldeq2}) and (\ref{dV})--(\ref{EE1}) that relate these functions to the scalars in (\ref{locscals}). However, it can be a difficult system to use in numerical work because time and radial derivatives cannot be (in general) decoupled. In the following section we will introduce an alternative set of covariant scalars whose corresponding evolution equations (see section XI) are wholly equivalent to (\ref{ev_13})--(\ref{cHam_13}) but are easier to handle.

\section{Quasi--local scalar functions and fluctuations.} 

The function $M$ that appears in (\ref{fieldeq1})--(\ref{fieldeq2}) is for LTB models the Misner--Sharp quasi--local mass--energy function, which is an important invariant in spherically symmetric spacetimes~\cite{MSQLM,hayward1,hayward2}. It is basically the integral along a spherical comoving domain of the field equation (\ref{fieldeq2}), which will be well defined if we assume the existence of a symmetry center (at $r=0$) and can be given as 
\ba 2M =\frac{2G}{c^2}\int_{\DD(r)}{\rho\,\FF\,\dd\VV_p}=\kappa\int_0^r{\rho R^2 R'\dd x},\label{MSmass}\\
\FF \equiv \sqrt{1+E}=\left[\dot R^2+1-\frac{2M}{R}\right]^{1/2}.\label{Fdef}\ea
where the spherical comoving domain $\DD(r)$ is defined further below and we have used the notation $\int_0^r{... \dd x}=\int_{x=0}^{x=r}{... \dd x}$. It is evident from (\ref{MSmass}) and (\ref{Fdef}) that $M$ is a proper volume integral ``weighed'' by the invariant scalar $\FF$ which generalizes the ``$\gamma$'' factor of Special Relativity \cite{hayward1,hayward2}. The integral definition (\ref{MSmass})  motivates the introduction of similar ``weighed'' proper volume integral functions which turn out to be very useful in the study of LTB models. In order to define such integrals, we remark that every regular $\T[t]$ contains comoving regions (containing a symmetry center) that are diffeomorphic to the product manifold    
\begin{equation}\DD(r)=\vartheta(r)\times \mathbb{S}^2(\theta,\phi),\qquad
 \vartheta(r) \equiv \{x\,|\, 0\leq x\leq r\},\label{etadef} \end{equation}  
where $x=0$ marks a symmetry center. Because of spherical symmetry every scalar function on a domain $\DD(r)$ is equivalent to a real valued function on the real interval $\vartheta(r)$.    

\subsection{Quasi--local scalar variables.}

\noindent
Consider now the following\\


\noindent\underline{Definition 1:  Quasi--local scalar functions $A_q$ }.\,\, Let $\vartheta(r)$ be a radial domain in a regular $\T[t]$. For every scalar function $A[t_0]:\vartheta(r)\to\mathbb{R}$, there is a ``quasi--local dual function'' $A_q[t_0]:\vartheta(r)\to\mathbb{R}$ such that
\begin{equation}  A_q[t_0](r)=\frac{\int_0^{r}{A\,\FF\,\dd\VV_p}}{\int_0^{r_0}{\FF\,\dd\VV_p}}=\frac{\int_0^{r}{A(t_0,x) R^2(t_0,x)R'(t_0,x)\,\dd x}}{\int_0^{r}{R^2(t_0,x)R'(t_0,x)\dd x}},\label{aveq_def}\end{equation}
where $\FF$ is given by (\ref{Fdef}). Since it is clear that $t$ is an arbitrary but fixed parameter in these integrals, we will omit (unless it is necessary) to express explicitly the time dependence of scalars $A$ and $A_q$.\\


\noindent\underline{Comment 1}:\, Definition (\ref{aveq_def}) is invariant under arbitrary rescalings of the radial coordinate $r=r(\bar r)$ that do not violate (\ref{noshx}) and (\ref{RirF}). In particular, the proper length $\ell=\int{\sqrt{g_{rr}}\dd r}$ can be used as the integration variable at each $\T[t]$ (though it is not useful as a global integral variable because $\ell=\ell(t,r)$). 

\smallskip

\noindent\underline{Comment 2}:\,  In general, not all hypersurfaces $\T[t]$ in LTB models are fully regular, since the central singularities (expanding or collapsing) are not simultaneous and will intersect some of the $\T[t]$.  For these $\T[t]$ the domain (\ref{etadef}) must be suitably restricted and the integrals in (\ref{aveq_def}) must be treated as improper integrals. We discuss this issue in \ref{restrictions}. 

\smallskip

\noindent\underline{Comment 3.} The quasi--local functions $A_q$ in (\ref{aveq_def}) are not ``averages'', as they do not comply with the properties of average distributions of a continuous random variable. They can be recast as proper volume averages with ``weight factor'' $\FF$ if we define them as functionals. See \cite{LTBave3} for discussion and clarification of this issue.

\subsection{Properties.}

The definition (\ref{aveq_def}) leads to the definition of a ``quasi--local volume'' given by
\begin{equation} \VV_q(r) = \int_{\DD(r)}{\FF\,\dd\VV_p}=4\pi\int_0^r{R^2R'\dd x}=\frac{4\pi}{3}R^3(r),\end{equation}
so that
\begin{equation} \frac{\dot\VV_q}{\VV_q}=\frac{3\dot R}{R}=\Theta_q,\qquad
\frac{\VV_q'}{\VV_q}=\frac{3 R'}{R},\label{dVq}\end{equation}
where we expressed (\ref{Theta1}) as $\Theta=[\ln(R^2R')]\,\dot{}$ and the commutation of $\partial/\partial t$ with the integrals in (\ref{aveq_def}). The quasi--local functions comply with the following properties:
\bse\ba    
 A_q'=(A_q)' = \frac{\VV_q'}{\VV_q}\,\left[A-A_q\,\right],\label{propq2}\\
A(r) - A_q(r) = \frac{1}{\VV_q(r)}\int_0^r{A' \,\VV_q \,\dd x},\label{propq3}\\
\dot A_q =(A_q)\,\dot{} =(\dot A)_q+(\Theta A)_q-\Theta_q A_q,\label{propq4}\ea\ese 

\subsection{A representation of covariant quasi--local scalars.}     

The introduced quasi--local variables provide a complete and very useful representation of covariant scalars that is alternative to (\ref{locscals}). 
Applying the definition (\ref{aveq_def}) to (\ref{fieldeq1}), (\ref{3Ricci1}) and (\ref{MSmass}) we obtain the quasi--local duals of $\rho,\,\RR$ and $\Theta$  
\begin{equation} 2m_q =\frac{2M}{R^3},\quad
k_q = -\frac{E}{R^2},\quad
\HH_q^2=\frac{\dot R^2}{R^2}=2m_q-k_q.\label{mkHq}\end{equation}
where, to simplify notation, we have defined
\begin{equation} 2m\equiv \frac{\kappa}{3}\,\rho,\qquad k\equiv \frac{\RR}{6},\qquad \HH\equiv\frac{\Theta}{3},\label{mkHdefs}\end{equation}
Notice that $M,\,R,\,\dot R=u^a\nabla_a R$ and $\FF=\sqrt{1+E}$ are invariants in spherical symmetry~\cite{MSQLM,hayward1,hayward2}, hence $m_q,\,k_q$ and $\HH_q$ are covariant quantities. As a consequence of (\ref{mkHq}), the scalars $\Sigma$ and $\EE$ in (\ref{Sigma1}) and (\ref{EE1}), associated with the shear and electric Weyl tensors, become expressible as deviations or fluctuations of the local scalars $\HH,\,m$ (or $\Theta,\rho$) with respect to their quasi--local duals:
\begin{equation}  \Sigma = -\left(\HH-\HH_q\right),\qquad 
\EE = -\left(m-m_q\right).\label{SigmaEE2}\end{equation}
Since the shear and electric Weyl tensors are covariant objects, then these fluctuations are also covariant. 

\subsection{Fluctuations of quasi--local duals.} 

A convenient way to relate the local covariant scalars $A$ in (\ref{locscals}) and their quasi--local duals $A_q$ follows by introducing the relative deviations (fluctuations)
\begin{equation} \Da \equiv \frac{A-A_q}{A_q}=\frac{A'_q/A_q}{3R'/R} =\frac{1}{A_q(r)}\int_0^r{A'\,V_q\,\dd x},\label{Dadef}\end{equation}
where we used (\ref{propq2}) and (\ref{propq3}). Considering (\ref{SigmaEE2}) and (\ref{Dadef}), each scalar $A$ in (\ref{locscals}) can be uniquely expressed in terms of its associated $A_q$ and $\Da$: 
\ba  m &=& m_q\,\left[1+\Dm\right],\quad \HH=\HH_q\,\left[1+\Dh\right],\nonumber
\\ k&=& k_q\,\left[1+\Dk\right],\quad\Sigma = -\HH_q\,\Dh,\quad \EE=-m_q\,
\Dm.\nonumber\\\label{qltransf}\ea 
By applying the definition (\ref{Dadef}) to the Friedman--like equation for $H_q^2$ in  (\ref{mkHq}) we obtain the following important constraint 
\begin{equation} 2\HH_q^2 \Dh = 2m_q\,\Dm- k_q\,\Dk,\label{HdH}\end{equation}
while the relation between $\Dm$ and $\Dk$ and radial gradients of the conventional variables $M$ and $E$
\begin{equation} \frac{M'}{M}=\frac{3R'}{R}\left[1+\Dm\right],\qquad \frac{E'}{E}= \frac{3R'}{R}\left[\frac{2}{3}+\Dk\right],\label{MED}\end{equation}
follows readily from (\ref{mkHq}) and (\ref{Dadef}). 

Since $m,\,k,\,\HH$ and $m_q,\,k_q,\,\HH_q$ are all covariant scalars, their $\Da$ are also covariant. In fact, these fluctuations can be understood in the context of a covariant gauge invariant and non--linear perturbation formalism (see section XII and references \cite{sussQL,suss2009}). Hence, the set 
\begin{equation}\{A_q,\Da\}=\{m_q,\,k_q,\,\HH_q,\,\Dm,\,\Dk,\,\Dh \},\label{qlscals}\end{equation} 
constitutes a complete covariant scalar representation that is equivalent and alternative to (\ref{locscals}).

There are important regularity issues associated with variables (\ref{qlscals}), which we discuss in detail in Appendix A. For example, any one of the $\Da$ might diverge if $A_q\to 0$ under regular conditions (no violation of (\ref{noshx}) and finite curvature scalars). However, as we show in Appendix \ref{divergingDa} this does not lead to a curvature singularity, as the Riemann tensor frame components do not diverge (see Appendix \ref{Riemann}). Also, the regular radial range of some of the hypersurfaces $\T[t]$ is necessarily restricted by a central singularity, which does not invalidate the regularity of these scalars away from the singular coordinate surface (see Appendix \ref{restrictions}). While these issues must be taken into consideration, they do not affect the usage of the variables (\ref{qlscals}).  

\section{Some theoretical and practical implications of quasi--local scalars.}

The quasi--local variables and their relative fluctuations introduced in the previous section lead in a natural way to an initial value parametrization of LTB models, which can be very helpful for numeric and qualitative work. Various elements of this parametrization has been already considered in the literature \cite{LTBchin,LTBfin,num2,num3}, but only as a useful set of coordinate ansatzes whose justification is merely their resemblance with FLRW parameters. However, these ansatzes can acquire a clear covariant meaning by its relation with quasi--local scalars.

\subsection{A FLRW--like metric and scaling laws for covariant scalars.}       

Consider the dimensionless scale factor 
\begin{equation} L\equiv\frac{R}{R_i},\label{Ldef}\end{equation}
where $R_i\equiv R(t_i,r)$ and $t=t_i$ marks a fiducial initial hypersurface $\T_i=\T[t_i]$. We shall denote henceforth all quantities evaluated at $t=t_i$ by the subindex ${}_i$. 
Bearing in mind (\ref{Ldef}), we can re--write (\ref{mkHq}) as 
\ba m_q &=& \frac{m_{qi}}{L^3},\quad k_q=\frac{k_{qi}}{L^2},\label{mkHL1}\\
\HH_q^2 &=& \frac{\dot L^2}{L^2}=2m_q-k_q=\frac{2m_{qi}-k_{qi}L}{L^3},\label{mkHL2}\ea
which, if we identify $L$ with a position dependent FLRW scale factor, are formally identical to the scaling laws for density, spatial curvature and Hubble factor in FRLR dust universes.  Applying this parametrization to the LTB metric (\ref{LTB1}) yields the FLRW--like line element
\begin{equation} \dd s^2=-c^2\dd t^2+L^2\left[\frac{\Gamma^2\,{R'_i}^2\,\dd r^2}{1-k_{qi}\,R_i^2}+R_i^2\left(\dd\theta^2+\sin^2\theta\dd \phi^2\right)\right],\label{LTB2}\end{equation}
where the new dimensionless metric function $\Gamma$ is 
\begin{equation} \Gamma \equiv \frac{R'/R}{R'_i/R_i}=1+\frac{L'/L}{R'_i/R_i},\qquad \Gamma_i=1. \label{Gammadef}\end{equation}
The local density ($m=\kappa\rho/3$) and spatial curvature ($k=\RR/6$) satisfy the following scaling laws in terms of $L,\,\Gamma$ and initial value functions
\bse\label{slaw}\ba  m &=& \frac{m_{qi}}{L^3}\,[1+\Dm] =\frac{m_i}{L^3\,\Gamma},\label{slaw1}\\
    k &+& \frac{k_{qi}}{L^2}\,[1+\Dk] = \frac{k_i}{L^2\,\Gamma}\,\left[1+\frac{\Gamma-1}{3\,(1+\Dik)}\right].\nonumber\\\label{slaw2}\ea\ese
where we used (\ref{fieldeq2}), (\ref{3Ricci1}) and (\ref{qltransf}). Comparing (\ref{slaw})  with (\ref{mkHL1}), and bearing in mind (\ref{mkHdefs}) and (\ref{Dadef}), we obtain the following scaling laws for $\Dm$ and $\Dk$
\bse\label{slawD}\ba 1+\Dm &=& \frac{1+\Dim}{\Gamma},\label{slawDm}\\
\frac{2}{3}+\Dk &=& \frac{2/3+\Dik}{\Gamma},\label{slawDk} \\
\Dm-\frac{3}{2}\Dk &=&\frac{\Dim-(3/2)\Dik}{\Gamma}.\label{slawD32}\ea\ese 
Applying (\ref{Dadef}) to the Hamiltonian constraint (\ref{mkHL2}) yields the scaling law for $\Dh$
\ba  2\Dh = \frac{2m_q\,\Dm-k_q\,\Dk}{2m_q-k_q}=\frac{2m_{qi}\,\Dm-k_{qi}\,L\,\Dk}{2m_{qi}-k_{qi}\,L}.\nonumber\\\label{slawDh}\ea
Inserting (\ref{slawDh}) into (\ref{SigmaEE2}) allows us to find  scaling laws similar to the ones above for $\HH=3\Theta$ and the scalars $\Sigma,\,\EE$, respectively associated by (\ref{Sigma1}) and (\ref{EE1}) to the shear and electric Weyl tensor. Using (\ref{Dadef}), we can express the scaling laws (\ref{slawD}) as  relations between the gradients $\HH'_q,\,m'_q$ and $k'_q$ and $L'$. Notice that it is always possible to use (\ref{mkHL2}) and (\ref{slawDh}) to eliminate any one of the three pairs $\HH_q,\,\Dh$ or $m_q,\,\Dm$ or $k_q$ and $\Dk$ in terms of the other two.

The scaling laws (\ref{slaw}), (\ref{slawD}), and (\ref{slawDh}) depend on both $L$ and $\Gamma$. While the effects of their dependence on $L$ is easy to grasp because the qualitative time dependence of $L$ (for a fixed $r$) can be appreciated directly from the Friedman--like equation (\ref{mkHL2}), there is no simple way to guess the qualitative behavior of $\Gamma$ (either in the $t$ or $r$ directions). As a consequence, there is not much we can do with these scaling laws as long as we lack an expression for $\Gamma$ in terms of $L$ and the initial value functions (we obtain this expression in section VII).

\subsection{A radial coordinate gauge.} 

The metric of LTB models in the form (\ref{LTB2}) is also invariant under an arbitrary rescaling $r=r(\bar r)$. Since $L$ and $\Gamma$ are time dependent, it is natural when using (\ref{LTB2})  to identify this radial coordinate gauge freedom with the freedom to choose the initial value function $R_i(r)$. However, because of (\ref{signRr}) and (\ref{RirF}), the choice for a function $R_i$ is not completely arbitrary: it depends on the topology of the space slices $\T[t]$ (see \cite{suss02} and Appendix \ref{topology}). 

For open models  (or LTB regions in which (\ref{noshx}) allows for $R'>0$ to hold everywhere) $R_i$ can be any monotonously increasing function complying with $R_i(0)=0$  and $R'_i>0$ for all $r$. Evidently, the simplest choice in these cases is
\begin{equation} R_i = R_0\,r,\label{rgauge}\end{equation}     
where $R_0$ is an arbitrary constant characteristic length scale. The gauge (\ref{rgauge}) is a popular choice in the literature \cite{LTBchin,LTBfin,num2,num3}, not only due to its simplicity, but because setting $r=R_i/R_0$ allows us to regard radial dependence as a dependence on an fiducial value of an invariant quantity ($R$) that has a clear physical and  geometric meaning. Also, the choice of $R_0$ provides a physical length scale for the radial coordinate. 

However, other radial coordinate gauges are often used (for example setting $M$ as radial coordinate, as in \cite{KH1,KH2,KH3,KH4}). And, as pointed before, (\ref{rgauge}) cannot be used in cases where a regular zero of $R'$ occurs. For closed models, $R_i$ must be selected so that it vanishes at both centers of symmetry and $R'_i$ has a common same order zero with $\FF=\sqrt{1+E}$ (to avoid the surface layer singularity associated with (\ref{layer}), see \cite{ltbstuff2}). Moreover, even if we use the parametrization associated with (\ref{LTB2}), we are not forced to use the radial coordinate gauge for fixing $R_i$, it is still possible to keep this initial value function unspecified and use the gauge freedom to fix any one of the other initial value functions.

\section{The analytic solutions in terms of an initial value approach. }

We rephrase in this section the analytic solutions (\ref{par1})--(\ref{ellt1}) in terms of  $\{m_{qi},\,k_{qi}\}$, which are related to the free parameters $M$ and $E$ in (\ref{par1})--(\ref{ellt1}) by 
\begin{equation} M = m_{qi}\,R_i^3,\qquad E = -k_{qi}\, R_i^2.\label{ME} \end{equation}
 A third free function (the ``bang time'', $\tbb$) necessarily appears as an ``integration constant'' in both, the solutions of $\dot L$ in (\ref{mkHL2}) and those of $\dot R$ in (\ref{fieldeq1}). However, as we show below, if we rewrite (\ref{par1})--(\ref{ellt1}) in the context of an initial value problem associated with $m_{qi},\,k_{qi}$ and $L$ (considering  that $L_i=1$), the bang time $\tbb$ can be obtained as a function of $m_{qi},\,k_{qi}$. 

Of course, other combinations of initial value functions, such as $\{m_{qi},\,\HH_{qi}\}$ or $\{k_{qi},\,\HH_{qi}\}$, can also be considered, as (\ref{mkHL2}) relates the three available functions. Since the fluctuations follow from the gradients of these functions from (\ref{Dadef}),  any two of $\{m_{qi},\,k_{qi},\,\HH_{qi}\}$ forms an irreducible set of initial value functions. As we show in section XI, these initial value functions provide also the basic initial conditions for the evolution equations under a numeric approach. 

\subsection{Parabolic models or regions: $k_{qi}=0$}

We express $M$ and $R$ in terms of $m_{qi}$ and $L$ from (\ref{Ldef}) and (\ref{ME}) in (\ref{par1}), after re--arranging terms we get the following closed analytic expression for $L$
\begin{equation} L =\left[1+\frac{3}{2}\sqrt{2m_{qi}}\,c(t-t_i)\right]^{2/3},\label{par2}\end{equation}
where we are only considering expanding configurations ($L$ increases for $t>t_i$). The bang time follows by setting $L=0$ and $t=\tbb$ in (\ref{par2})
\begin{equation}  c\tbb = ct_i-\frac{2}{3\sqrt{2m_{qi}}}=ct_i-\frac{2}{3\HH_{qi}}.\label{tbbpar}  \end{equation}

\subsection{Hyperbolic models or regions: $k_{qi}< 0$}

We obtain the following implicit solution of the form $t=t(R)$ by eliminating the parameter $\eta$ from the equation for $R$ in (\ref{hypt1}) and substituting in the equation for $t$:
\begin{equation} \frac{E^{3/2}}{M}\,c(t-\tbb)=Z_h(\bar R),\label{hypZR}\end{equation}
where $\bar R= (E/M)R$ and $Z_h$ is the function 
\begin{equation} u  \mapsto Z_h(u)=u ^{1/2} \left( {2 + u } \right)^{1/2}  - \hbox{arccosh}(1 + u ).\label{hypZ1a}  \end{equation}
We express then $M,\,E$ and $R$ in (\ref{hypZR}) in terms of  $m_{qi},\,k_{qi}$ and $L$ from (\ref{Ldef}) and (\ref{ME}). The result is
\begin{equation} y_i\, c(t-t_i) = Z_h(x_iL) - Z_h(x_i),\label{phi_L_h}\end{equation}
where
\begin{equation}
 x_i = \frac{|k_{qi}|}{m_{qi}},\qquad y_i=\frac{|k_{qi}|^{3/2}}{m_{qi}}.\label{xy}
\end{equation}
Setting $t=\tbb$ and $L=0$ in (\ref{phi_L_h}) and using (\ref{hypZ1a}) yields the bang time as a function of $m_{qi}$ and $|k_{qi}|$:
\begin{equation} c\tbb = ct_i-\frac{Z_h(x_i)}{y_i}.\label{tbbhe} \end{equation}

\subsection{Elliptic models or regions: $k_{qi}> 0$}

In this case, the implicit solution $t=t(R)$ follows from eliminating $\eta$ from the first equation in (\ref{ellt1}) and substituting in the second one. The resulting implicit solution  has two branches, an ``expanding'' one ($0<\eta<\pi$ with $\dot R>0$) and a ``collapsing'' one ($\pi<\eta<2\pi$ with $\dot R<0$):
\begin{equation} \frac{|E|^{3/2}}{M}\,c(t-\tbb)=\left\{ \begin{array}{l}
  Z_e(\bar R)\qquad\quad\; \hbox{expanding phase} \\ 
  2\pi-Z_e(\bar R) \quad \hbox{collapsing phase} \\ 
 \end{array} \right.,\label{ellZR} \end{equation}
where $\bar R= (|E|/M)R$ and $Z_e$ is given by
\begin{equation} u  \mapsto  Z_e(u)= \arccos(1 - u )-u ^{1/2} \left( {2 - u } \right)^{1/2}.\label{ellZ1a}  \end{equation}
Proceeding as in the hyperbolic case, we transform (\ref{ellZR}) into
\begin{equation}  y_i\, c(t-t_i)+Z_e(x_i)  = \left\{ \begin{array}{l}
 Z_e(x_iL) \qquad\qquad \hbox{expanding phase}\\ 
  \\ 
 2\pi-Z_e(x_iL) \qquad \hbox{collapsing phase}\\ 
 \end{array} \right.\label{ellZ2}\end{equation}
where
\begin{equation}
 x_i = \frac{k_{qi}}{m_{qi}},\qquad y_i=\frac{k_{qi}^{3/2}}{m_{qi}},\label{xye}
\end{equation}
It follows from (\ref{ellZ2})--(\ref{ellZ1a}) that $L$ is restricted by $0<L\leq \Lmax$, where the maximal expansion is  $\Lmax=2/x_i=2m_{qi}/k_{qi}$, characterized by  $\dot L=0$ and $\HH_q=0$. 

Setting $t=\tbb$ and $L=0$ in the expanding phase of (\ref{ellZ2}) yields a bang time function, while $t=\tcoll$ and $L=0$ in the collapsing phase yields the ``crunch'' time associated with the collapsing singularity. The maximal expansion time follows by substituting  $t=\tmax$ and $L=\Lmax$ in either branch of (\ref{ellZ2}). These times are given by
\bse\label{tmc}\ba  c\tbb &=& ct_i-\frac{Z_e(x_i)}{y_i},\label{tmc1}\\
 c\tmax &=& c\tbb+\frac{\pi}{y_i}=ct_i+\frac{\pi-Z_e(x_i)}{y_i},\label{tmc2}\\  c\tcoll &=& c\tbb+\frac{2\pi}{y_i}=ct_i+\frac{2\pi-Z_e(x_i)}{y_i}\label{tmc3}.\ea\ese
Notice that (in general) $\tmax=\tmax(r)$ and $\tcoll=\tcoll(r)$, so neither one coincides with a $\T[t]$ hypersurface (like $\tbb(r)$). For every comoving observer $r =$ const., the time evolution is contained in the range $\tbb(r)<t<\tcoll(r)$.

Rewriting the analytic solutions (\ref{par1})--(\ref{ellt1}) as in (\ref{par2}), and as the implicit forms (\ref{phi_L_h}) and (\ref{ellZ2}) is very useful, as it allows us to relate (by implicit radial derivation) the gradients of $L$ (and thus of $R$) with gradients of the intial value functions $m_{iq},\,k_{iq}$ (see section 11).  Notice that the time splitting associated with $u^a$ and the scaling of $R$ with $R_i$ in $L$ imposes on LTB models a constraint between $\tbb$ and $m_{qi},\,k_{qi}$, since (\ref{tbbpar}) and (\ref{tbbhe}) imply that any choice of these functions uniquely determines the ``age'' $t_i-\tbb$ of the initial slice $t=t_i$ for any comoving observer. Also, the initial slice $t=t_i$ in elliptic models always occurs in the expanding phase~\cite{suss02}.

\section{Curvature singularities.}

The new variables allow us to express the coordinate locus of curvature singularities {\it {exclusively}} in terms of $L$ and $\Gamma$ by means of (\ref{slaw}), (\ref{slawD}) and (\ref{slawDh}).  Notice that for reasonable initial value functions (bounded and continuous), all scalars $A_q=m_q,\,k_q,\,\HH_q$ diverge as $L\to 0$, whereas local scalars $A=m,\,k,\,\HH$ can also diverge if $\Gamma\to 0$ (even if $L>0$). Following the standard criteria that define a curvature singularity and considering (\ref{Riemann2}), we can identify two known possible singular surfaces
\bse\ba L(t,r) &=& 0,\qquad \hbox{central singularity}\label{Lzero}\\
 \Gamma(t,r) &=& 0 \qquad \hbox{shell crossing singularity}.\label{Gzero}\ea\ese
Notice that if $\Gamma>0$, then all scalars $A$ and $A_q$ only diverge at the central singularity $L=0$, but if $\Gamma\to 0$, then all the relative fluctuations $\Da$ diverge (with $A_q\ne 0$), so that local scalars $A$ diverge while their quasi--local duals $A_q$ remain bounded. This is an obviously unphysical effect of shell crossings that must be avoided.  

It is important to remark that characterizing the curvature singularities as in (\ref{Lzero}) and (\ref{Gzero}) represents an improvement in clarity over the conventional variables. In these latter variables ``$R=0$'' can be either a regular symmetry center, or a curvature singularity. Likewise, ``$R'=0$'' can be either a shell crossing singularity or a regular ``turning value'' $r=\rtv$ associated with a closed model. There is no such ambiguity in our parametrization, since $L> 0$ holds at a symmetry center and $\Gamma>0$ holds at the turning value.  

In order to test the avoidance of (\ref{Gzero}) we need to compute $\Gamma$. For this purpose, we derive both sides of the solutions (\ref{par2}), (\ref{phi_L_h}) and (\ref{ellZ2}), then we use (\ref{Dadef}) and $L'/L=(1-\Gamma)R_i'/R_i$ to eliminate $m'_{qi},k'_{qi}$ and $L'$ in terms of $\Dim,\,\Dik$ and $\Gamma$.  The result is

\begin{itemize}
\item {\underline{Parabolic models or regions.}}
\begin{equation} \Gamma = 1+\Dim-\frac{\Dim}{L^{3/2}}.\label{Gp}\end{equation}
\item {\underline{Hyperbolic and elliptic models or regions.}}
\ba  \Gamma = 1+3(\Dim-\Dik)\left(1-\frac{\HH_q}{\HH_{qi}}\right)\nonumber\\
-3\HH_q\,c(t-t_i)\,\left(\Dim-\frac{3}{2}\Dik\right),\label{Ghe}\ea
where $\HH_q$ and $\HH_{qi}$ follow from (\ref{mkHL2}), while $c(t-t_i)$ is given by (\ref{phi_L_h}) and (\ref{ellZ2}).
\end{itemize}

\noindent
Inserting (\ref{Gp}) or (\ref{Ghe}) (depending on the kinematic class) into (\ref{slaw}), (\ref{slawD}) and (\ref{slawDh}) leads to closed analytic expressions for all these scaling laws in terms of $L$ and initial value functions in the context of the initial value approach described in previous sections.   The functional forms of $\Gamma$ and $\HH_q$ in (\ref{mkHL2}) and (\ref{Gp}) allow us to obtain analytic expressions for all the scaling laws of covariant scalars (as functions of $L$ and initial value functions).  

From (\ref{noshx}), (\ref{Dadef}), (\ref{Gammadef}) and (\ref{slawDm}) applied to $A=m$,  condition (\ref{noshx}) for standard regularity that avoids (\ref{Gzero}) now reads
\begin{equation} \Gamma > 0 \quad\forall\;\; (ct,r)\quad\hbox{such that}\quad L>0 \label{noshxG}\end{equation}
The fulfillment of this condition is examined in the following section.

\section{The Hellaby--Lake conditions.} 

We obtain in this section the Hellaby--Lake conditions (\ref{noshx1})--(\ref{noshx2}) that fulfill the regularity condition (\ref{noshxG}) \cite{ltbstuff,suss02,ltbstuff1}, but now as restrictions on the initial value functions. These conditions guarantee absence of shell crossings for all $t$. Notice from (\ref{noshx}) and (\ref{Gammadef}) that avoidance of this singularity already excludes a surface layer singularity associated with (\ref{layer}).

In order to examine the Hellaby--Lake conditions in the initial value parametrization, we need to relate the gradients $M'$ and $E'$ appearing in (\ref{noshx1})--(\ref{noshx2}) with the initial fluctuations by specializing (\ref{MED}) to $t=t_i$
\begin{equation} \frac{M'}{M}=\frac{3R'_i}{R_i}\left[1+\Dim\right],\qquad \frac{E'}{E}=\frac{3R'_i}{R_i}\left[\frac{2}{3}+\Dik\right].\label{MEDi}\end{equation}
We will also need an expression for the radial gradient $\tbb'$ in terms of our initial value functions. This expression follows by differentiating both sides of (\ref{tbbpar}) and (\ref{tbbhe}) with respect to $r$:
\bse\ba  \frac{c\tbb'}{3R'_i/R_i} &=& c(t_i-\tbb)\,\Dim =\frac{2\,\Dim}{3\sqrt{2m_{qi}}},\qquad \hbox{parabolic}\nonumber\\\label{tbbrp}\\ 
 \frac{c\tbb'}{3R'_i/R_i} 
&=& \frac{\Dim-\Dik}{\HH_{qi}}-c(t_i-\tbb)\left(\Dim-\frac{3}{2}\Dik\right),\nonumber\\&{}& \hbox{hyperbolic and elliptic}\label{tbbr}\ea\ese
where $\tbb$ in (\ref{tbbr}) is given by (\ref{tbbhe}) with $Z_h(x_i)$ or $Z_e(x_i)$, respectively, for hyperbolic and elliptic models. As mentioned before, $R_i$ can be always be prescribed as a radial coordinate gauge.

\subsection{Parabolic models/regions.}

From (\ref{Gp}), if $L\to 0$ then $\Gamma\to-\Dim/L^{3/2}$, while $\Gamma\to 1+\Dim$ for $L\to\infty$. Considering (\ref{slawDm}), the necessary and sufficient condition for (\ref{noshxG}) to hold is simply 
\begin{equation} -1\leq \Dim\leq 0,\label{noshx_par}\end{equation}      
which from (\ref{Dadef}) implies $m'_{qi}\leq 0$. It is evident from (\ref{MEDi}) and (\ref{tbbrp}) that (\ref{noshx_par}) implies $\tbb'/R'_i\leq 0$ and $M'\geq 0$, and so it is equivalent  to the Hellaby--Lake conditions (\ref{noshx1}). 

\subsection{Hyperbolic models/regions.}

For a fixed $r$ all initial value functions are finite, consider then $t_i>t_1>\tbb$ but $t_1\approx \tbb$, so that $L\approx 0$ and $\HH_q\to\infty$, and thus (\ref{Ghe}) becomes 
\ba \Gamma &\approx& -\frac{3\HH_q}{\HH_{qi}}\left[\Dim-\Dik-\HH_{qi}\,c(t_i-t_1)\left(\Dim-\frac{3}{2}\Dik\right)\right]\nonumber\\
&=& -3\HH_q\left[\frac{c\tbb'}{3R'_i/R_i}+c(t_1-\tbb)\,\left(\Dim-\frac{3}{2}\Dik\right)\right],\nonumber\\ \label{Gh1}\ea
where we used (\ref{tbbr}). Now consider $t\to\infty$ so that $L\to\infty$ for all $r$. From (\ref{mkHL2}) and (\ref{phi_L_h}) we have in this limit $\HH_q\to 0$ and
\begin{equation} \HH_q c(t-t_i)\approx 1+\frac{Z_h(x_i)+\ln(2x_i L)-2}{x_iL}+O(L^{-2}),\label{Hqthyp}\end{equation}
hence $\Gamma$ in (\ref{Ghe}) becomes
\begin{equation} \Gamma\approx 1+\frac{3}{2}\Dik+O(L^{-2}).\label{Gh2}\end{equation}
Bearing in mind, from (\ref{noshx}) and (\ref{signRr}), that $R'_i>0$ for hyperbolic models/regions and that (\ref{noshxG}) must hold as $t_1\to\tbb$ in (\ref{Gh1}), and since the scaling law (\ref{slawDm}) and (\ref{Gh1})--(\ref{Gh2}) must hold for all $t$ and $r$, the conditions for (\ref{noshxG}) are \footnote{This equation corrects an error in \cite{suss02}, where $-2/3\leq \Dik\leq 0$ was incorrectly assumed.} 
\begin{equation} c\tbb'\leq 0,\qquad \Dik\geq -\frac{2}{3},\qquad \Dim\geq -1.\label{noshxGh}\end{equation}
where $\tbb'$ is given in terms of our initial value functions by (\ref{tbbr}). Notice, from (\ref{slawDk}), that all regular hyperbolic models comply with $\Dk\geq -2/3$  for all their time evolution. From (\ref{MED}) and (\ref{MEDi}), it is evident that (\ref{noshxGh}) are completely equivalent to the Hellaby--Lake conditions (\ref{noshx1}). Since these conditions are necessary and sufficient, (\ref{noshxGh}) are also necessary and sufficient~\cite{ltbstuff,suss02,ltbstuff1}.

\subsection{Elliptic models/regions.}

We look again at $\Gamma$ in (\ref{Ghe}), but now considering that $\HH_q$ changes sign (from positive to negative) at some $t=\tmax$ where $L=L_{\rm{max}}=2m_{qi}/k_{qi}=2/x_i$, so that $0<L\leq L_{\rm{max}}$ and we have the presence of a second (collapsing) singularity as $L\to 0$ when $t\to\tcoll$, with $\tmax$ and $\tcoll$ given by (\ref{tmc}). Near the maximal expansion $L\to L_{\rm{max}}$ and so $\HH_q\to 0$, while $c(t-t_i)\to c(\tmax-t_i)$. Thus, we have in this limit the following necessary (not sufficient) condition for (\ref{noshxG}): 
\begin{equation} \Gamma= 1+3(\Dim-\Dik)>0.\label{Gmax}  \end{equation}
Near the initial singularity (bang) we have $\HH_q\to\infty$ as $t\to\tbb$, while near the collapsing singularity we have $\HH_q\to-\infty$ as $t\to\tcoll$ with $\tcoll$ given by (\ref{tmc}). We consider $t=t_1$ such that $t_i>t_1>\tbb$ but $t_1\approx\tbb$,  as well as $t=t_2$ such that $\tcoll>t_2>t_i$ with $t_2\approx \tcoll$. For these times $\Gamma$ takes the form
\ba \Gamma &\approx& -\frac{3\HH_q}{\HH_{qi}}\left[\Dim-\Dik-\HH_{qi}\,c(t_i-t_1)\left(\Dim-\frac{3}{2}\Dik\right)\right]\nonumber\\
&=& -3\HH_q\left[\frac{c\tbb'}{3R'_i/R_i}+c(t_1-\tbb)\,\left(\Dim-\frac{3}{2}\Dik\right)\right],\nonumber\\ \label{Ge1}\ea
\ba \Gamma &\approx& 3\frac{|\HH_q|}{\HH_{qi}}\left[\HH_{qi}\,c(t_2-t_i)\left(\Dim-\frac{3}{2}\Dik\right)+\Dim-\Dik\right]\nonumber\\
&=& 3|\HH_q|\left[\frac{c\tbb'}{3R'_i/R_i}+c(t_2-\tbb)\,\left(\Dim-\frac{3}{2}\Dik\right)\right], \nonumber\\ \label{Ge2}\ea
Since  (\ref{Ge1}) must hold as $t_1\to\tbb$ and (\ref{Ge2}) must hold as $t_2\to \tcoll$, each expression furnishes a second and third necessary condition for (\ref{noshxG}). Since both conditions that follow from (\ref{Ge1}) and (\ref{Ge2}) must hold for all $t$ in the range $\tbb(r)<t<\tcoll(r)$ for all $r$, these two conditions taken together are then the necessary and sufficient conditions for the fulfillment of (\ref{noshxG}), which can be written as
\begin{equation} \frac{c\tbb'}{3R'_i/R_i}\leq 0,\qquad  \frac{c\tcoll'}{3R'_i/R_i}\geq 0,\qquad \Dim\geq -1,\label{noshxGe}\end{equation}
with
\begin{equation} \frac{c\tcoll'}{3R'_i/R_i} =\left(\Dim-\frac{3}{2}\Dik\right)\,c(\tcoll-\tbb)+\frac{c\tbb'}{3R'_i/R_i},\label{tcollr}\end{equation}
where we used $c(\tcoll-t_i)=c(\tcoll-\tbb)-c(t_i-\tbb)$, with $\tbb$ and $\tcoll$ given by (\ref{tmc}), and the equal sign holds only at a symmetry center. The fact that one of the three conditions in (\ref{noshxGe}) to avoid shell crossings in elliptic models is basically a sign condition on the gradient $\tcoll'$ has not been, apparently, noticed in the extensive literature that uses the conventional variables. The reader is requested to compare the simplicity and elegance of (\ref{noshxGe}) with the cumbersome form (\ref{noshx2}).

Conditions (\ref{noshxGe}) imply the following necessary (but not sufficient) condition for (\ref{noshxG})
\begin{equation}\Dim-\frac{3}{2}\Dik\geq 0,\label{noshx_nec1}\end{equation}
which, because of (\ref{slawD32}), implies that $\Dm-(3/2)\Dk\geq 0$ necessarily holds for all times if (\ref{noshxG}) holds. Since condition (\ref{noshx_nec1}) implies (\ref{Gmax}), then the latter (also necessary but not sufficient) follows from (\ref{noshxGe}) too.   

It is straightforward to show from (\ref{Dadef}), (\ref{tmc}), (\ref{MEDi}), (\ref{tbbr}) and (\ref{tcollr}) that conditions (\ref{noshxGe}) are equivalent to the Hellaby--Lake conditions (\ref{noshx2}). We have expressed these conditions in terms of $m_{qi},\,k_{qi},\,\Dim,\,\Dik$, but it is straightforward to rewrite them in terms of other sets initial value functions, like $m_{qi},\,\HH_{qi},\,\Dim,\,\Dih$. Because of (\ref{signRr}) and (\ref{Gammadef}), we do not need to express (\ref{noshxGe}) in terms of the $\pm$ signs for $R'$, as it is understood that the signs of $\tbb'$ and $\tcoll'$ must be, respectively, the same and the opposite of the sign of $R'_i$ (which because of (\ref{signRr}) will be the same as the sign of $R'$).

\section{Omega and Hubble expansion parameters.}

The fact that $\HH_q$ behaves in (\ref{mkHL2}) as a Hubble scalar of a FLRW dust universe, has motivated some authors \cite{LTBfin,num3} dealing with cosmological and astrophysical applications to define an ``Omega'' parameter analogous to that of a FLRW dust cosmology
\begin{equation} \hOm \equiv \frac{\kappa\rho_q}{3\HH_q^2}=\frac{2m_q}{\HH_q^2}=\frac{2m_q}{2m_q-k_q}=\frac{2m_{qi}}{2m_{qi}-k_{qi}L},\label{Omdef1}\end{equation}
so that 
\begin{equation} \hOm-1 = \frac{k_q}{2m_q-k_q} =\frac{k_{qi}L}{2m_{qi}-k_{qi}L},\label{Omdef2}\end{equation}
where we use the symbol $\hOm$, and not $\Omega_q$, because this quantity is not the quasi--local dual of the scalar $2m/\HH^2$ under the definition (\ref{aveq_def}) (see section XIII). The following FLRW--like scaling laws hold for $\hOm$ and $\hOm-1$:
\begin{equation} \hOm=\frac{\hOmi}{\hOmi-(\hOmi-1)\,L},\qquad \hOm-1=\frac{(\hOmi-1)L}{\hOmi-(\hOmi-1)\,L},\label{Omdef3}\end{equation}
while the Hamiltonian constraint (\ref{mkHL2}) can be rewritten as an analogue of the FLRW Hubble expansion 
\begin{equation} \HH_q^2 = \HH_{qi}^2\left[\frac{\hOmi}{L^3}+\frac{1-\hOmi}{L^2}\right].\label{OmHdef} \end{equation}
The relative fluctuation (\ref{slawDh}) of the quasi--local Hubble factor, $\Dh$,  takes the appealing form
\begin{equation} 2\Dh = \hOm\,\Dm + (\hOm-1)\,\Dk,\label{DhOm}\end{equation}
which can be given in terms of metric functions $L,\,\Gamma$ and initial value functions $\hOmi,\,\Dim$ and $\Dih$ by means of (\ref{slawDm}), (\ref{slawDk}), (\ref{slawDh}) and (\ref{Omdef3}). 

We can identify $\hOmi$ in (\ref{Omdef3}), (\ref{OmHdef}) and (\ref{DhOm}) as analogous to an $r-$dependent FLRW Omega factor at a fiducial cosmic time $t_i$ 
\begin{equation} \hOmi=\frac{2m_{qi}}{\HH_{qi}^2}=\frac{2m_{qi}}{2m_{qi}-k_{qi}},\qquad \hOmi-1=\frac{k_{qi}}{\HH_{qi}^2}=\frac{k_{qi}}{2m_{qi}-k_{qi}},\label{Omidef}\end{equation}
Notice that the sign of $\hOmi-1$ in any region of $r$ determines the sign of $\hOm-1$ for all times, which is the same as with $k_{qi}$ and $k_q$. Hence, $\hOm-1$ behaves as the FLRW  Omega factor for spatial curvature, so that $\hOmi-1$ determines the kinematic class: parabolic if $\hOmi=1$, elliptic if $\hOmi>1$ and hyperbolic if $0<\hOmi<1$. 
Also, irrespective of the kinematic class we have for every LTB model or region:\, $\hOm\to 1$ as $L\to 0$ (near the central curvature singularity). For all hyperbolic models or regions $\hOm$ is bounded between 0 and 1 for all choices of $\hOmi,\,\HH_{qi}$ (or  $m_{qi},\,k_{qi}$), hence:\, $\hOm\to 0$ as $L\to\infty$, whereas for elliptic models or regions $\hOm\to \infty$ as $L\to \Lmax$ because $\HH_q\to 0$.

\subsection{Analytic solutions in terms of $\hOmi$}

It is straightforward to parametrize the analytic solutions in terms of $\hOmi,\,\HH_{qi}$. For the parabolic case we have $\hOmi=1$, and thus we just obtain (\ref{par2}) with $\sqrt{2m_{qi}}=\HH_{qi}$. For the hyperbolic and elliptic models or regions we substitute
\begin{equation} m_{qi} = \HH_{qi}^2\,\hOmi,\qquad k_{qi}=\HH_{qi}^2\,(\hOmi-1),\label{mkOm}\end{equation}
in (\ref{phi_L_h}) and (\ref{ellZ2}). Since $\hOmi,\,\HH_{qi}$ are roughly equivalent to inhomogeneous generalization of fiducial Omega and Hubble factors,  parametrizing the analytic solutions with these initial value functions could be more intuitive than doing it with the density and spatial curvature profiles $m_{qi},\,k_{qi}$.  The solutions (\ref{phi_L_h}) and (\ref{ellZ2}) become  

\begin{itemize}
\item
{\underline{Hyperbolic models or regions: $\hOmi-1\leq 0$}.}
\begin{equation} c(t-t_i)=\frac{W-W_i}{\HH_{qi}}.\label{hypZ3} \end{equation}
\item
{\underline{Elliptic models or regions: $\hOmi-1\geq 0$}.}
\begin{equation} \HH_{qi}\,c(t - t_i )  = \left\{ \begin{array}{l}
 W-W_i \qquad\hbox{expanding phase}\\ 
  \\ 
 \pi\hOmi(\hOmi-1)^{-3/2} - W-W_i\\ \qquad \hbox{collapsing phase}\\ 
 \end{array} \right.\label{ellZ3}\end{equation}
\end{itemize}
where the functions $W$ and $W_i$ are
\begin{widetext}
\noindent Hyperbolic models or regions
\bse\ba 
     W &=& \frac{\left[\hOmi+(1-\hOmi)L\right]^{1/2}L^{1/2}}{1-\hOmi}-\frac{\hOmi}{2(1-\hOmi)^{3/2}}\hbox{arccosh}\left(\frac{2L}{\hOmi}+1-2L\right),\label{hypZ4a}\\
    W_i &=& \frac{1}{1-\hOmi}-\frac{\hOmi}{2(1-\hOmi)^{3/2}}\hbox{arccosh}\left(\frac{2}{\hOmi}-1\right).\label{hypZ4b}\ea\ese
\noindent Elliptic models or regions
\bse\ba 
     W &=& \frac{\hOmi}{2(\hOmi-1)^{3/2}}\hbox{arccos}\left(\frac{2L}{\hOmi}+1-2L\right)-\frac{\left[\hOmi-(\hOmi-1)L\right]^{1/2}L^{1/2}}{\hOmi-1},\label{ellZ4a}\\
    W_i &=& \frac{\hOmi}{2(\hOmi-1)^{3/2}}\hbox{arccos}\left(\frac{2}{\hOmi}-1\right)-\frac{1}{\hOmi-1}.\label{ellZ4b}\ea\ese
\end{widetext}

\noindent
Setting $t=\tbb$ and $L=0$ in (\ref{hypZ3}) and in the expanding phase of (\ref{ellZ3}) yields the bang time for hyperbolic and elliptic configurations
\begin{equation}  c\tbb = ct_i-\frac{W_i}{\HH_{qi}},\label{tbbOm}\end{equation}
while the maximal expansion and collapse times, $t=\tmax$ and $t= \tcoll$, given by (\ref{tmc}) are
\bse\label{tmcOm}\ba\tmax=c\tbb+\frac{\pi\hOmi}{2 \HH_{qi} [\hOmi-1]^{3/2}},\\  
\tcoll=c\tbb+\frac{\pi\hOmi}{\HH_{qi} [\hOmi-1]^{3/2}}.\ea\ese

We can also parametrize $\Gamma$, as well as the Hellaby--Lake conditions, in terms of $\hOmi,\,\HH_{qi}$ and their gradients. For the parabolic case, we substitute into (\ref{Gp}) and (\ref{noshx_par})
\begin{equation}\Dim = \frac{2\HH'_{qi}/\HH_{qi}}{3R'_i/R_i},\end{equation}
while for hyperbolic and elliptic models or regions we substitute into (\ref{Ghe}), (\ref{noshxGh}) and (\ref{noshxGe}) equation (\ref{mkOm}) plus
\bse\label{OmD}\ba
\Dim &=& \frac{R_i}{3R_i'}\left[\frac{\hOmi'}{\hOmi}+\frac{2\HH_{qi}'}{\HH_{qi}}\right],\\ \Dik &=& \frac{R_i}{3R_i'}\left[\frac{\hOmi'}{\hOmi-1}+\frac{2\HH_{qi}'}{\HH_{qi}}\right],\ea\ese
where $R_i$ can be specified as a choice of radial coordinate.

The initial value functions $\hOmi,\,\HH_{qi}$ can also be used in the analytic solutions given in the forms (\ref{par1})--(\ref{ellt1}), in terms of the parameter $\eta$. For this purpose, we eliminate $M$ and $E$ by means of 
\begin{equation} \hOmi=\frac{2M}{2M+E\,R_i},\quad \HH_{qi}^2=\frac{2M+E\,R_i}{R_i^3},\label{OmME}\end{equation}
where $R_i$ can be prescribed as a radial coordinate gauge. The bang time is then given by (\ref{tbbOm}). 

\section{Mapping of density to density or velocity to velocity profiles. } 

Krasi\'nski and Hellaby \cite{KH1,KH4} have examined an interesting proposal: the possibility that arbitrary radial profiles of density (or velocity or density and velocity) given at two different cosmic times for a finite radial range can be fit (or ``mapped'') by a unique LTB model. The initial value parametrization of the analytic solutions is ideally suited to examine this issue, which can also be understood in terms of the equivalence of boundary and initial conditions to solve the evolution equations of sections XI and XIII. Our aim here is to illustrate how the initial value parametrization can be helpful, hence we only examine in detail the mapping between the quasi--local density profiles by a hyperbolic model, since all other similar mappings can be handled in the same manner. 

Consider two radial profiles of $m_q$. Since $t=t_i$ is arbitrary, then $m_{qi}$ can be one of the profiles, while we take $m_{qj}=m_q(ct_j,r)$ for some $t_j>t_i$ as the second profile. If $m_{qi}$ and $m_{qj}$ are known, then
\begin{equation}L_j(r) = L(ct_j, r) =\left(\frac{m_{qi}}{m_{qj}}\right)^{1/3}\label{Lij}\end{equation}
follows from (\ref{mkHL1}) (remember that $L_i=1$). For a parabolic model, we have from (\ref{par2})      
\begin{equation} c(t_j-t_i) = \frac{2}{3\sqrt{2m_{qi}}}\left[\left(\frac{m_{qi}}{m_{qj}}\right)^{1/2}-1\right],\end{equation}
a constraint that will not be satisfied unless $m_{qi}$ and $m_{qj}$ have very restricted forms. For a hyperbolic model we rewrite (\ref{phi_L_h}) as
\begin{equation} F(|k_{qi}|)\equiv \frac{Z_h(\alpha|k_{qi}|)-Z_h(\beta|k_{qi}|)}{\beta|k_{qi}|^{3/2}}=c(t_j-t_i), \label{densdens1}\end{equation}
where $Z_h$ is given by (\ref{hypZ1a}) and
\begin{equation} \alpha =\frac{1}{m_{qi}^{2/3}m_{qj}^{1/3}},\qquad \beta=\frac{1}{m_{qi}}.\end{equation}
Since a unique LTB model is determined by prescribing $m_{qi}$ and $|k_{qi}|$, and we assume $m_{qi}$ and $m_{qj}$ known, then (\ref{densdens1}) becomes a constraint to find the missing initial value function $|k_{qi}|$ that determines the LTB model. Solving this constraint for known $m_{qi},\,m_{qj}$ and $c(t_j-t_i)$ can only be done numerically, but by looking at the properties of the function $F(|k_{qi}|)$ we can find the conditions for the existence of this solution, as it was done in \cite{KH1,KH4}.

Considering that in an expanding model we have $m_{qi}>m_{qj}$ for $t_j>t_i$, we have $\alpha>\beta$ in general, and since $Z_h$ is monotonously increasing, then the left hand side of (\ref{densdens1}) is positive (like the right hand side). We also have
\bse\ba F(|k_{qi}|) \approx \frac{\sqrt{2}}{3\beta}\left(\alpha^{3/2}-\beta^{3/2}\right)+O(|k_{qi}|),\qquad |k_{qi}|\approx 0,\nonumber\\
\\
F(|k_{qi}|) \approx \frac{\alpha-\beta}{\beta|k_{qi}|}+O(|k_{qi}|^{-3/2}),\qquad |k_{qi}|\to \infty,\nonumber\\\ea\ese
which, together with similar expansions of $dF/d|k_{qi}|$, show that $F(|k_{qi}|)$ is monotonously decreasing and bounded by
\begin{equation}  0< F(|k_{qi}|) \leq \frac{\sqrt{2}}{3\beta}\left(\alpha^{3/2}-\beta^{3/2}\right)=\frac{\sqrt{2}}{3}\left[m_{qj}^{-1/2}-m_{qi}^{-1/2}\right]\end{equation}
Hence, considering (\ref{Lij}), a unique solution of (\ref{densdens1}) must exist if
\begin{equation}  1+\frac{3}{2}\sqrt{2m_{qi}}c(t_j-t_i)<\left(\frac{m_{qi}}{m_{qj}}\right)^{1/2}=L_j^{3/2},\label{densdens2} \end{equation}
which is the same result found in \cite{KH1}. Comparison with (\ref{par2}) shows that the density-density mapping requires the LTB model that maps $m_{qi}$ to $m_{qj}$ to expand faster than a parabolic model. For the elliptic case we follow the same method: using the solution (\ref{ellZ2}) (in the expanding and collapsing phases) as a constraint to find the missing initial value function $k_{qi}$. If the local density profiles $m_i$ and $m_j$ are known (instead of $m_{qi}$ and $m_{qj}$), then we have to use the scaling law (\ref{slaw1}) with $\Gamma$ given by (\ref{Gammadef})
\begin{equation} \frac{3m_i}{m_j}R_i^2R'_i =3\,L_j\,\Gamma_j\,R_i^2R'_i=(R_i^3\,L_j^3)\,',\end{equation}
which yields $L_j$ by integration (as $m_i$ and $m_j$ are known and $R_i$ can be fixed as a coordinate gauge). It is straightforward to show that the analytic solutions for the hyperbolic class yield a unique LTB model mapping these profiles if (\ref{densdens2}) holds.

Krasi\'nski and Hellaby also examined the mapping of profiles of velocities defined as $v\equiv \dot R/R_i =L\HH_q$. To deal with this case we eliminate $k_{qi}$ as  $k_{qi}=2m_{qi}-\HH_{qi}^2$, so that the initial value functions are $m_{qi}$ and $\HH_{qi}$. Then, from the Hamiltonian constraint (\ref{mkHL2}) we obtain
\begin{equation} L_j =\frac{2m_{qi} R_i^2}{v_j^2-v_i^2+2m_{qi} R_i^2},\end{equation}
which inserted in the solutions (\ref{phi_L_h}) and (\ref{ellZ2}) yields the constraints needed to obtain the missing initial value function $m_{qi}$. We can also obtain the LTB model that results from assuming that the profiles $\hOmi$ and $\hOm_j$ are known. In this case we obtain $L_j$ from (\ref{Omdef3})
\begin{equation} L_j = \frac{\hOmi}{\hOm_j}\frac{1-\hOm_j}{1-\hOmi},\label{LjOm}\end{equation}
which substituted in the solutions (\ref{hypZ3}) and (\ref{ellZ3}) yields the missing initial value function $\HH_{qi}$ by means of the constraint
\begin{equation} \HH_{qi} = \frac{W(\hOmi,\hOm_j)-W_i(\hOmi)}{c(t_j-t_i)},\label{OmOm}\end{equation}
where $W$ and $W_i$ take the forms (\ref{hypZ4a})--(\ref{hypZ4b}) or (\ref{ellZ4a})--(\ref{ellZ4b}) with $L$ given by (\ref{LjOm}) for hyperbolic or elliptic models or regions. Evidently, obtaining the missing initial value function by means of constraints like (\ref{densdens1}) and (\ref{OmOm}) does not guarantee the fulfillment of regularity conditions (like absence of shell crossings). This has to be verified independently by means of the Hellaby--Lake conditions once the LTB model mapping the profiles has been found.  

The issue of mapping two profiles can also be understood in terms of the systems of  evolution equations that we derive in the following section. These systems are partial differential equations, hence initial and boundary conditions are given as functions of $r$ (or radial profiles). The selection of two initial value functions is sufficient to determine a unique solution. These functions can be, either $m_{qi},\HH_{qi}$ or $m_{qi},k_{qi}$ or $\hOmi,\,\HH_{qi}$. Hence if we know only one of these functions (evaluated at $t=t_i$) and the same function at $t=t_j$, then we may use one of the available constraints to find the second initial value function, and thus guarantee (if the constraint has a solution) that the evolution system yields a unique LTB model. We have proceeded in this section (following \cite{KH1, KH4}) by using the analytic solutions as the constraints that link the boundary and initial conditions through $L$. The results are the same as those of \cite{KH1, KH4}, but the initial value parametrization of the solutions makes the whole procedure a lot more natural and straightforward.

\section{Evolution equations for a numerical treatment.}

The analytic solutions, either in terms of the conventional variables $M,\,E,\,c\tbb,\,R$ or the initial value parametrization $m_{qi},\,k_{qi},\,R_i,\,L$ that we have presented, are mostly useful for qualitative work (as illustrated in the previous section). We remark that these solutions are either parametric or implicit, and thus a pure analytic framework based on them is necessary limited. For concrete applications in models of less idealized cosmological inhomogeneities and observations there is no other way but to consider a numeric framework. 

The 1+3 fluid flow system (\ref{ev_13})--(\ref{cHam_13}) discussed in section IV provides a fully covariant description and can be used to determine the dynamics of LTB models by means of numeric techniques. However, the associated spacelike constraints (\ref{c_13}) are partial differential equation on $r$ that are not easy to solve, making it difficult (in general) to decouple the time and radial derivatives. As we show in this section, the quasi--local scalars and their fluctuations (in the initial value parametrization) provide more convenient variables for a numeric treatment of LTB models. 

\subsection{Quasi--local evolution equations.} 

An alternative system to the evolution equations (\ref{ev_13})--(\ref{cHam_13}) follows by considering a 1+3 fluid flow evolution equations for the quasi--local scalars (\ref{qlscals}), involving $m_q,\,\HH_q$ and their corresponding  relative fluctuations $\Dm,\,\Dh$. This system can be obtained by eliminating $\rho,\,\Theta,\,\Sigma$ and $\EE$ in (\ref{ev_13}) in terms of the new variables by means of (\ref{qltransf}) and using (\ref{Dadef}) to deal with the constraints (\ref{c_13}). Considering the notation (\ref{mkHdefs}): $\HH=\Theta/3,\,\, m=\kappa\rho/3$ and $k=\RR/6$, this framework leads to the following evolution equations
\bse\label{EVql}\ba \dot m_q &=& -3 m_q\HH_q,\label{EVql1}\\
\dot \HH_q &=& -\HH_q^2-m_q, \label{EVql2}\\
\dot\delta^{(m)} &=& -3(1+\Dm)\,\HH_q\Dh,\label{EVql3}\\
\dot\delta^{(\HH)} &=& -(1+\Dh)\,\HH_q\Dh+\frac{m_q}{\HH_q}(\Dh-\Dm),\nonumber\\\label{EVql4}\ea\ese
while the spacelike constrains (\ref{c_13}) become the equations defining $\Dm$ and $\Dh$ through (\ref{propq2}) and (\ref{Dadef}), and the Hamiltonian constraint (\ref{cHam_13}) reduces to (\ref{mkHL2}). This is the particular case of pure dust in \cite{suss2009,sussQL}. 

The system (\ref{EVql}) is equivalent to (\ref{ev_13}), but it is much more practical and easier to use in a numerical treatment, as it is not necessary to solve any radial differential equation as a precondition to solve it. In practice, we can handle it as a system of non--linear autonomous ordinary differential equations, where $r$ enters as a parameter (see Appendix B of \cite{suss08}). Notice that once we have solved (\ref{EVql}) all local scalars in (\ref{locscals}) can be determined by means of (\ref{qltransf}), (\ref{mkHL2}) and (\ref{slawDh}).

Since $\Dh$ diverges as $\HH_q\to 0$, as the maximal expansion $ct=c\tmax$ is reached in elliptic models or regions (see \ref{divergingDa}), it is better to use in these cases the alternative variable $\Sigma=-\HH_q\Dh$ (see (\ref{qltransf})). This leads to the alternative system
\bse\label{EVqlsig}\ba \dot m_q  &=& -3 m_q\HH_q,\label{EVql11}\\
\dot \HH_q &=& -\HH_q^2-m_q, \label{EVql22}\\
\dot\delta^{(m)} &=& -3(1+\Dm)\,\Sigma,\label{EVql33}\\
\dot\Sigma &=& -2\,\Sigma\,\HH_q+\Sigma^2+m_q\,\Dm,\label{EVql44}\ea\ese
which is entirely equivalent to (\ref{EVql}), and has the advantage (besides its use for recollapsing configurations) that $\Sigma$ is the scalar associated with the shear tensor by (\ref{Sigma1}).

We may supplement either (\ref{EVql}) or (\ref{EVqlsig}) with the following extra pair of differential equations 
\bse\label{EVql56}\ba  \dot L &=& L\,\HH_q,\label{EVql5}\\
\dot\Gamma &=& 3\,\Gamma \,\HH_q\Dh = -3\,\Gamma\,\Sigma,\label{EVql6}\ea\ese
which follow directly from (\ref{qltransf}), (\ref{mkHL2}) and (\ref{Gammadef}). By including these extra equations we can use the systems above to determine numerically the metric functions in (\ref{LTB2}). In fact, if we substitute the  scaling laws (\ref{mkHL1})--(\ref{mkHL2}), (\ref{slaw}), (\ref{slawD}) and (\ref{slawDh}) into (\ref{EVql56}) we can work only with these two evolution equations given as
\bse\label{EVql5566}\ba  
\dot L &=& \frac{[2m_{qi}-k_{qi}L]^{1/2}}{L^{1/2}},\label{EVql55}\\
\dot \Gamma &=& -\frac{3}{2}\,\frac{2m_{qi}\,[\Dim+1-\Gamma]-k_{qi}L\,[\Dik+\frac{2}{3}(1-\Gamma)]}{[2m_{qi}-k_{qi}L]^{1/2}},\nonumber\\\label{EVql66}\ea\ese
whose solution ($L$ and $\Gamma$) allows us to compute all scalars by means of these scaling laws. However, from a numeric point of view it might be more convenient and practical to work with (\ref{EVql}) or (\ref{EVqlsig}) supplemented by (\ref{EVql56}) than with (\ref{EVql5566}).    

Solving any one of the systems of evolution equations described above is needed to find radial null geodesics and their corresponding red shift factor, $z$, which are in turn  needed for computing observable quantities (luminosity distance, etc). Given a numerical solution of either system, radial geodesics $ct=ct_N(r)$ in the past null cone and $z$ follow from solving \cite{kras2}
\bse\label{null}\ba \frac{c\,\dd t_N(r)}{\dd r} &=& -\frac{R_N'}{(1+E)^{1/2}}=-\frac{L_N\,\Gamma_N\,R'_i}{[1-k_{qi}R_i^2]^{1/2}},\nonumber\\\label{radgeo}\\
 \frac{\dd}{\dd r}\ln (1+z) &=& \frac{\dot R_N'}{(1+E)^{1/2}}\nonumber\\
 &=&\frac{L_N\,\Gamma_N\,\HH_{qN}\,(1+3\Dh_N)\,R'_i}{[1-k_{qi}R_i^2]^{1/2}},\label{redshift}\ea\ese
where the subindex ${}_N$ means evaluation at the null curve $ct=ct_N(r)$, {\it{i.e.}} $A_N=A(ct_N(r),r)$. Since the systems (\ref{EVql}) or (\ref{EVqlsig}) or (\ref{EVql5566}) can be handled in practice as constrained ordinary differential equations, solving numerically (\ref{null}) can be achieved with simple numerical techniques. 

\subsection{Initial conditions.}   

An obvious choice of initial value functions for the systems (\ref{EVql})--(\ref{EVql56}) or (\ref{EVqlsig})--(\ref{EVql56}) is
\ba  m_{qi},\quad\HH_{qi},\quad \Dim = \frac{m'_{qi}/m_{qi}}{3R'_i/R_i},\nonumber\\
 \Dih = \frac{\HH'_{qi}/\HH_{qi}}{3R'_i/R_i}\quad\hbox{or}\quad \Sigma_i = -\frac{\HH'_{qi}}{3R'_i/R_i}\label{initconds1} \ea
where we used (\ref{Dadef}) and $R_i$ can be prescribed as a radial coordinate gauge.        

Since $k_{qi}$ determines the solutions of (\ref{mkHL2}) (or (\ref{fieldeq1}) for $R=R_iL$), this initial value function is closely related to the kinematic evolution of the models (or specific regions of them). Hence, it is more practical and intuitive to choose instead of (\ref{initconds1}) the initial value functions  
\begin{equation} m_{qi},\qquad k_{qi},\qquad \Dim = \frac{m'_{qi}/m_{qi}}{3R'_i/R_i},\qquad \Dik = \frac{k'_{qi}/k_{qi}}{3R'_i/R_i}.\label{initconds2} \end{equation}
These initial conditions are sufficient to solve (\ref{EVql5566}), but to solve (\ref{EVql})--(\ref{EVql56}) or (\ref{EVql56})--(\ref{EVqlsig}) we also need  
$\HH_{qi}$ and $\Dih$ or $\Sigma_i$, which readily follow from (\ref{mkHL2}) and (\ref{slawDh})
\ba \HH_{qi} &=& [2m_{qi}-k_{qi}]^{1/2},\nonumber\\
 \Dih &=& \frac{2m_{qi}\Dim-k_{qi}\Dik}{2[2m_{qi}-k_{qi}]}\nonumber\\
 &\hbox{or}&\quad \Sigma_i=-\frac{2m_{qi}\Dim-k_{qi}\Dik}{2[2m_{qi}-k_{qi}]^{1/2}},\nonumber\\\label{initconds3}\ea
The main advantage of using (\ref{initconds2}) as initial conditions is the fact that they are exactly the same initial value functions that we employed in the parametrization of the analytic solutions and in the forms of $\Gamma$ in (\ref{Gp})--(\ref{Ghe}), and the Hellaby--Lake conditions. Hence, given a choice of these functions, a time evolution free from shell crossings can be immediately tested by means of (\ref{noshx_par}), (\ref{noshxGh}) and (\ref{noshxGe}). 

The evolution equations (\ref{EVqlsig}) and have already been used for a dynamical systems approach to LTB models \cite{suss08}, while a suitable generalization of (\ref{EVql}) was employed for a numeric study of cosmological models endowed with the LTB metric but with an energy--momentum tensor of an anisotropic fluid~\cite{suss2009}.  

A comprehensive numeric study of LTB models by means of the systems (\ref{EVql}),\, (\ref{EVqlsig}) or (\ref{EVql5566}) is beyond the scope of this article, and thus is presently under examination in a separate work. Our purpose has been to show how these systems, constructed with quasi--local variables, are potentially promising for this purpose.

\section{Gauge invariant and covariant spherical perturbations on a FLRW background.}

The connection between the system (\ref{EVql}) and a perturbation formalism on a FLRW background is clearly suggested by the fact that the evolution equations (\ref{EVql1}) and (\ref{EVql2}) for the quasi--local scalars $m_q,\,\HH_q$ are formally identical to the energy balance and Raychaudhuri equations of a dust FLRW cosmology. Also, from (\ref{mkHq}), (\ref{mkHdefs}), (\ref{mkHL1}) and (\ref{mkHL2}), we have $\dot k_q=-2k_q \HH_q$, which is the same evolution equation for $\RR$ in a FLRW spacetime, while the $\Da$ defined by (\ref{Dadef}) and its relation with the scalars in (\ref{qltransf}) clearly highlights a sort of perturbation definition. We provide in this section a rigorous characterization of this intuitive resemblance.   

A perturbation formalism between an idealized spacetime, $\bar S$ (a dust FLRW cosmology), and an ``lumpy'' universe model, $S$ (a dust LTB model), follows by defining a ``background model'' in $S$, constructed by objects (in $S$) that are the images of a suitable map $\Phi$ between objects in $\bar S$ to objects in $S$~\cite{ellisbruni89}. Though, perturbations in general are not uniquely defined by such an abstract map $\Phi$, because of the ambiguity of the gauge freedom associated with choices of coordinates and hypersurfaces. This fact requires carefully defining perturbed variables that are ``gauge--invariant''~\cite{bardeen}. However, in the specific case that concerns us we face a simpler task, as both the idealized FLRW and the perturbed LTB spacetimes are LRS (locally rotationally symmetric) and can be completely described by scalar functions~\cite{LRS}. Hence, $\Phi$ can be a map of scalars to scalars, while the fact that both spacetimes are spherically symmetric and are given in the same normal geodesic coordinate (or frame) representation, greatly suppresses most of the gauge freedom that we would find for a general $S$.\footnote{The coordinate resemblance between FLRW dust cosmologies and LTB models is further highlighted by using metric form (\ref{LTB2}) for the latter. }  

The map $\Phi$ can be defined rigorously as follows. Let $\bar X$ and $X$ be, respectively, the sets of smooth integrable scalar functions in $\bar S$ and $S$, then for all covariant FLRW  scalars $\bar A\in \bar X$ (we denote FLRW objects with an over--bar) the map
\begin{equation} \Phi: \bar X\to X,\qquad \bar A \mapsto\Phi(\bar A)=A_q\in X,\label{Phi}\end{equation}
defines a ``background model'' (associated to a FLRW cosmology) in LTB models through the quasi--local scalars $A_q$ (which are LTB objects satisfying FLRW dynamics). Given their common normal geodesic representation and considering the perturbations
\begin{equation} \Da = \frac{A-\Phi(\bar A)}{\Phi(\bar A)}\label{DPhi}\end{equation} 
associated with (\ref{Phi}), LTB models characterized by the metric (\ref{LTB1}) and source (\ref{Tab}) can be considered then as spherical non--linear ``perturbed'' dust FLRW cosmologies in the  ``comoving'' gauge.   

The perturbation scheme that we described fits naturally to the gauge invariant and covariant (GIC) approach of Dunsbury, Ellis and Bruni~\cite{ellisbruni89,BDE}. Following these authors, a perturbation scheme on FLRW cosmologies is covariant if the ``lumpy'' model $S$ is described by variables defined in the framework of the 1+3 fluid flow variables associated with the system (\ref{ev_13}). Although our description of LTB spacetimes is not based on these scalars (the representation (\ref{locscals})) but on the quasi--local representation (\ref{qlscals}), it is still a covariant description because $m_q$ and $\HH_q$, are covariant scalars by virtue of their connection with the invariants $M,\,R$ and their derivatives in (\ref{mkHq}). Hence, the formalism based on $m_q$ and $\HH_q$ would also be covariant. 

As commented by Ellis and Bruni~\cite{ellisbruni89}, by virtue of the Stewart--Walker gauge invariance lemma \cite{SWlemma}, all covariant objects in $S$ that would vanish in the background $\bar S$ (a FLRW cosmology in this case) are gauge invariant (GI), to all orders, and also in the usual sense (as in \cite{bardeen}). The tensorial quantities in LTB models that vanish for a dust FLRW cosmology in the 1+3 formalism are the shear and electric Weyl tensors $\sigma^{ab}$ and $E^{ab}$, given by (\ref{Sigma1}) and (\ref{EE1}) in terms of the scalar functions $\Sigma$ and $\EE$, which from (\ref{SigmaEE2}), can be written in terms of the fluctuations $m-m_q$ and $\HH-\HH_q$. From (\ref{Dadef}), it is evident that these fluctuations and $\Da$, as well as the radial gradients $m_q'$ and $\HH_q'$  are all ``first order'' quantities (in the perturbation scheme) that are GI to all orders, though  $m_q,\,\HH_q$ are not (which is expected because these are ``zero order'' background variables). As a consequence, the fluid flow dynamics of LTB models in the quasi--local scalar representation (\ref{qlscals}) also provides a rigorous characterization of LTB models as spherical, non--linear GIC perturbations on a FLRW background.               

We note that spatial curvature $k=\RR/6$ and its quasi--local dual $k_q=\RR_q/6$ do not appear in the dynamic equations (\ref{EVql1})--(\ref{EVql4}), though they can be used when specifying initial conditions as in (\ref{initconds2}) (see also the appendices of \cite{suss2009}). Spatial curvature is GI only if the FLRW cosmology $\bar S$ is spatially flat, but its associated variables $\Dk$ and $k_q'$ are GI. In fact, from (\ref{HdH}), (\ref{mkHL1}) and (\ref{mkHL2}) we can always eliminate either one of $m_q,\,\HH_q$ or $\Dm,\,\Dh$ in terms of $k_q$ and $\Dk$, and construct a system of evolution equations equivalent to (\ref{EVql1})--(\ref{EVql4}), but describing the dynamics in terms of spatial curvature  $k_q$ and its perturbation $\Dk$.

Given the correspondence between (\ref{EVql}) and the dynamics of non--linear perturbations on a FLRW background, it is important to examine the connection with linear perturbation theory of dust sources. For this purpose we derive a second order equation for the density perturbation $\Dm$ by differentiating both sides of (\ref{EVql1}) and use the remaining equations (\ref{EVql2})--(\ref{EVql4}) to
eliminate all derivatives except $\ddot\delta^{(\mu)}$ and $\dDm$. We obtain
\ba  \ddot\delta^{(m)} &-&\frac{[\dDm]^2}{1+\Dm}+2\HH_q\,
\dDm-\frac{\kappa}{2}\,\rho_q\,\Dm\,\left(1+\Dm\right)=0,\nonumber\\\label{Dm2_eveq}\ea
which is an exact non--linear equation for $\Dm$ (a similar equation was obtained in \cite{kasai}). For near homogeneous conditions, as assumed in linear perturbations with ``small'' perturbations $|\Dm|\ll 1$, (\ref{Dm2_eveq}) reduces to
\begin{equation}\ddot\delta^{(m)}+2\HH_q\,
\dDm-\frac{\kappa}{2}\,\rho_q\,\Dm=0, \label{lp_iso}\end{equation}
an equation that is formally identical to the evolution equation for linear density perturbations of a dust source around a FLRW background (characterized by $\rho_q,\,\HH_q$) in the comoving gauge~\cite{padma}, which for dust is a synchronous gauge as well.

\section{A theoretical context for the Omega parameter.}

The parameter $\hOm$ that we discussed in section IX has been introduced in the literature \cite{LTBfin,num3} (together with $\HH_q$) as useful ansatzes justified by their resemblance to the corresponding FLRW Hubble and Omega parameters. Another generalization of the FLRW Hubble factor, suggested by Moffat and Tartarsky \cite{moftar} and used in various articles \cite{num2}, follows by defining two expansion factors, a radial and an azimuthal one, which can be given in terms of $\HH_q$ and $\Dh$ as:
\bse\ba H_\perp &\equiv& \frac{\dot R}{R}=\frac{\dot L}{L}=\HH_q,\\
H_\parallel &\equiv&  \frac{\dot R'}{R'}=\frac{\dot L}{L}+\frac{\dot \Gamma}{\Gamma}=\HH_q-3\Sigma=\HH_q(1+3\Dh),\nonumber\\\ea\ese 
Evidently, $H_\perp$ and $H_\parallel$ correspond, respectively, to expansion factors of proper lengths in the direction orthogonal and parallel to radial rays orthogonal to the orbits of SO(3). Moffat and Tartarsky define from these expansion factors an ``effective'' Hubble parameter as 
\begin{equation}H_{\tiny{\textrm{eff}}}^2=H_\perp^2+2 H_\perp H_\parallel=\HH_q\HH=\HH_q^2(1+\Dh),\end{equation}
so that an Omega parameter follows as $\Omega_{\tiny{\textrm{eff}}}=\kappa\rho/(3H_{\tiny{\textrm{eff}}}^2)$. However, both pairs $\HH_q,\,\hOm$ and $H_{\tiny{\textrm{eff}}},\,\Omega_{\tiny{\textrm{eff}}}$ are basically useful quantities in the qualitative or numeric application of LTB models, since the proper local covariant generalization of the FLRW Hubble parameter to LTB models is neither $\HH_q$, nor $H_{\tiny{\textrm{eff}}}$, but the expression given by equation (41) of \cite{HMM}. Nevertheless, it is still interesting to discuss the theoretical assumptions underlying these expressions. 

While both $\HH_q$ and $H_{\tiny{\textrm{eff}}}$ are covariant quantities, the theoretical context for $\HH_q$ may be easier to justify, as this scalar is the quasi--local dual of the fluid flow expansion scalar $\HH=\Theta/3$, and thus it is a GIC background variable in the perturbation formalism discussed in section XII (it is the image of the FLRW Hubble parameter under the map (\ref{Phi})). However, $\HH_q^2$ is not the quasi--local dual of $\HH^2$, since the definition (\ref{aveq_def}) implies that $(\HH^2)_q\ne (\HH_q)^2$, hence $\HH_q^2$ (and  thus) $\hOm$ are not images under the perturbation map (\ref{Phi}) of the Omega and squared Hubble parameters of a FRLW dust spacetime. As a consequence $\hOm$ is not a background variable in the perturbation formalism, though its gradient defined as
\begin{equation}\Delta \equiv \frac{\hOm'/\hOm}{3R'/R}=\Dm-2\Dh = (1-\hOm)\,(\Dm-\Dk),\end{equation}
is a GIC perturbation in this formalism (since $\Dm$ and $\Dh$ are), though we have $\Delta\ne (\Omega-\hOm)/\hOm$ in general (this relation only holds in the linear limit). On the other hand, it is hard to find a theoretical context for $H_{\tiny{\textrm{eff}}},\,\Omega_{\tiny{\textrm{eff}}}$ besides their resemblance to FLRW quantities and their utility in computations.   

\subsection{Evolution equations in terms of $\hOm$ and $\Delta$.}

Since  $\hOm$ and $\HH_q$ in (\ref{Omdef1}) and (\ref{OmHdef}) are constructed from $m_q$ and $k_q$, both quantities can easily be computed from a numeric solution of the systems (\ref{EVql}) or (\ref{EVqlsig}). The initial value functions $m_{qi}$ and $k_{qi}$ in (\ref{initconds1}), (\ref{initconds2}) and (\ref{initconds3}) can be given in terms of $\hOmi$ and $\HH_{qi}$ from (\ref{mkOm}), while $\Dim$ and $\Dik$ follow from (\ref{OmD}).

However, given the practical utility of $\hOm$ and the connection of its gradient $\Delta$ to the perturbation formalism of section XII, it is still useful to construct a system of evolution equations that include evolution laws for these quantities. Rewriting the system (\ref{EVql}) in terms of $\hOm$ and $\Delta$ yields the following  dimensionless system
\bse\label{EVqlOm}\ba \frac{\partial\textrm{H}}{\partial\tau} &=& -\textrm{H}^2\,\left(1+\frac{1}{2}\hOm\right),\label{EVql111}\\
\frac{\partial\hOm}{\partial\tau}&=&\textrm{H}\,\hOm\,\left(\hOm-1\right),\label{EVql222}\\
\frac{\partial\Delta}{\partial\tau} &=&\textrm{H}\,\left[\left(\Dh+\Delta\right)\hOm-\left(1+3\Delta+4\Dh\right)\,\Dh\right],\nonumber\\\label{EVql333}\\
\frac{\partial\Dh}{\partial\tau} &=& -\textrm{H}\,\left[\left(1+\Dh\right)\Dh+\frac{1}{2}\left(\Dh+\Delta\right)\hOm\right],\nonumber\\\label{EVql444}\ea\ese
where we have introduced the dimensionless variables
\begin{equation} 
\textrm{H} \equiv \frac{\HH_q}{H_0},\qquad
\tau \equiv H_0\,c(t-t_i),\label{dless}\end{equation}
with $H_0$ an inverse length scale ($\textrm{cm}^{-1}$), which can be identified as the   cosmological Hubble scale at a specific fiducial cosmic time (we can also identify $R_0=H_0^{-1}$ when specifying the radial coordinate gauge through $R_i$). The initial conditions for this system are basically $\HH_{qi},\,\hOmi$ and their radial gradients, which makes their  specification very intuitive and practical, as these initial value functions could tend at a given asymptotic limit to the Hubble and Omega factors in a FLRW background at a fiducial cosmic time. The system (\ref{EVqlOm}) can also be supplemented by the differential equations (\ref{EVql56}) for $L$ and $\Gamma$. 

\subsection{A dynamical system}

The system (\ref{EVqlOm})  clearly suggests a dynamical systems approach, as discussed in \cite{DS1} for FLRW and Bianchi models. This approach is based on the fluid flow equations \cite{ellisbruni89,BDE,1plus3} and an Omega parameter constructed as the ratio of density to the the squared expansion scalar $\HH^2=\Theta^2/9$. As shown in \cite{DS1}, if the evolution parameter is defined by $\partial/\partial\xi =(1/\HH)\partial/\partial \tau$ the Raychaudhuri equation decouples from the remaining evolution equations, leading to a reduced system that can be analyzed qualitatively: critical points, invariant subspaces, etc. For LTB models associated with evolution equations like (\ref{EVqlOm}), it is more convenient (see \cite{suss08}) to define $\xi$ in terms of $\textrm{H}$ in (\ref{dless}) by means of the coordinate transformation \cite{suss08}
\begin{equation} \tau =\tau(\xi,\bar r),\qquad r=\bar r,\qquad \xi=\ln\,\textrm{H}\end{equation}
so that for every scalar $A(\tau,r)=A(\tau(\xi,r),r)=A(\xi,r)$ and derivatives associated with the 4--velocity flow $\partial/\partial \tau$ (with $r$ constant) become
\ba  \left[\frac{\partial A}{\partial\tau}\right]_r=\frac{\partial A}{\partial\xi}\,\left[\frac{\partial \xi}{\partial\tau}\right]_r=\frac{\partial A}{\partial\xi}\,\textrm{H}\nonumber\\
\Rightarrow\quad \frac{\partial}{\partial\xi}=\frac{1}{\textrm{H}}\frac{\partial}{\partial\tau}
=\frac{1}{\HH_q}\frac{\partial}{c\partial t}.\ea
Hence, the three equations (\ref{EVql222})--(\ref{EVql444}) become independent of $\textrm{H}$ and effectively decouple from (\ref{EVql111}), leading to the reduced system
\bse \label{DS}\ba \frac{\partial\hOm}{\partial\xi} &=& \hOm\,\left(\hOm-1\right),\label{DS1}\\
\frac{\partial\Dh}{\partial\xi} &=&-\left(1+\Dh\right)\Dh-\frac{1}{2}\left(\Dh+\Delta\right)\hOm,\label{DS3}\\
\frac{\partial\Delta}{\partial\xi} &=&\left(\Dh+\Delta\right)\hOm-\left(1+3\Delta+4\Dh\right)\,\Dh,\nonumber\\\label{DS2}\ea\ese
which is very similar to that analyzed in \cite{suss08} (where $\Dm$ and a dimensionless $\Sigma$ were used instead of $\Delta$ and $\Dh$). Notice that this dynamical system is characterized by a  3--dimensional phase space parametrized by $\{\hOm,\,\Delta,\,\Dh \}$, which contains the FLRW dust case as an invariant set given by the line $[\hOm=\hOm(\xi),\Delta=0,\,\Dh=0]$. Therefore, this dynamical systems approach also provides a nice and appealing theoretical context to justify the role of $\hOm$ as an inhomogeneous generalization of the FLRW dust Omega parameter.

\section{Regularity of special LTB configurations.} 

\subsection{Closed elliptic models and regular zeroes of $R'=0$.}

In closed elliptic models the $\T[t]$ are homeomorphic to $\mathbb{S}^3$ (see Appendix \ref{topology}). There are two symmetry centers, at $r=0$ and $r=r_c$. Since $R(t,0)=R(t,r_c)=0$ for all $t$, then $R'(\rtv)=0$ where $0<\rtv<r_c$ must hold for all $t$. The regularity conditions (\ref{signRr}) and (\ref{RirF}) imply that for all $r$ the sign of $R'$ is the same for all $\T[t]$. Hence, (\ref{noshx}) requires the function $R_i(r)$ to be selected so that $R_i(0)=R_i(r_c)=0$ and $R_i'(\rtv)=0$, with  
\ba R'_i>0\quad \hbox{and}\quad 0<\FF\leq 1\quad\;\;\;\; \hbox{for}\quad 0<r<\rtv,\nonumber\\
R'_i<0\quad \hbox{and}\quad -1\leq \FF<0\quad \hbox{for}\quad \rtv<r< r_c,\nonumber\\\label{S3top}\ea
where $\FF=\pm\sqrt{1+E}$, with $\FF(0)=1,\,\,\FF(r_c)=-1$ and $\FF'(0)=\FF'(r_c)=\FF(\rtv)=0$. Since $R,\,R'$ and $\FF$ are bounded, if (\ref{noshx}) holds then the maximal coordinate range is $0\leq r\leq r_c$ in all regular $\T[t]$.

Equations (\ref{S3top}) provide the coordinate ranges where the ``$+$'' or ``$-$'' signs hold in specifying the Hellaby--Lake conditions (\ref{noshx2}) in the conventional variables. However, the $\pm$ signs are no longer needed with the initial value parametrization, as it is clear that fulfillment of (\ref{noshxGe}) (the equivalent of (\ref{noshx2})) implies by its construction that $\tbb'$ and $\tcoll'$ must have, respectively, the opposite and same sign (and a common zero) as $R_i'/R_i$, and from (\ref{signRr}), this sign will be the same for $R'/R$ at all $\T[t]$.  

If only the first sign equality in (\ref{noshx}) holds, then there would exist a range of $r$ for which 
\begin{equation}\hbox{sign}(R'_i)=\hbox{sign}(M')\ne \hbox{sign}(\FF), \label{layer} \end{equation}
so that $M'$ and $R'_i$ have a common zero, but either $\FF$ has a zero that is not common to that zero, or $\FF$ has no zeroes. In these situations, $\rho$ in (\ref{fieldeq2}) remains bounded, and so there is no curvature singularity, but we have a surface layer discontinuity at $r=\rtv$~\cite{ltbstuff,suss02,ltbstuff2}. This is the reason why when $R'$ vanishes regularly at some $r=\rtv$, this worldline must lie, either in an elliptic model or in an elliptic region. Besides the surface layer, (\ref{layer}) also implies an ill--defined metric coefficient $g_{rr}$ in (\ref{LTB1}) and (\ref{LTB2}) (which leads in turn to an ill--defined proper radial length between different comoving dust layers becomes).

Once  these issues are taken under consideration, initial conditions complying with the Hellaby--Lake conditions (\ref{noshxGe}) can be provided for this class of models, whether for analytic/qualitive or numeric work.

\subsection{Simultaneous big--bang.} 

In general, the initial curvature singularity given by (\ref{Lzero}) is not simultaneous, since it is marked by the curve $ct=c\tbb(r)$ in the $(ct,r)$ coordinate plane. For parabolic models, the condition $c\tbb'=0$ in (\ref{tbbrp}) implies $\Dim=0$, and thus $m_{qi}'=0$, which corresponds to the FLRW limit. However, non--trivial hyperbolic and elliptic LTB models follow by setting the initial value functions so that $c\tbb$ is a constant (see \cite{KH4}). In order to examine these cases, we use (\ref{tbbhe}), (\ref{tmc}) and (\ref{tbbr}) to rewrite $\Gamma$ in (\ref{Ghe}) in terms of $c\tbb$ and $c\tbb'$, which after setting $\tbb'=0$ and $\tbb=\tbb^{(0)}=$ constant, leads to
\begin{equation} \Gamma =1+3(\Dim-\Dik)-3\HH_q\,c(t-\tbb^{(0)})\left(\Dim-\frac{3}{2}\Dik\right), \label{Gsimtbb} \end{equation}
where $\HH_q$ follows from (\ref{mkHL2}) and
\begin{equation} c(t-\tbb^{(0)})=\left\{ \begin{array}{l}
  Z_h(x_i\,L)/y_i\quad \hbox{hyperbolic} \\ 
  Z_e(x_i\,L)/y_i \quad \hbox{elliptic} \\ 
 \end{array} \right.,\label{ctZbb} \end{equation}
with $x_i,\,y_i,\,Z_h$ and $Z_e$ given by (\ref{xy}), (\ref{hypZ1a}), (\ref{xye}) and (\ref{ellZ1a}). Since $x_i$ and $y_i$ depend only on $m_{qi}$ and $k_{qi}$, a simultaneous big--bang $\tbb'=0$ implies  the following constraint on these initial value functions 
\begin{equation} c(t_i-\tbb^{(0)})=\left\{ \begin{array}{l}
  Z_h(x_i)/y_i\quad \hbox{hyperbolic} \\ 
  Z_e(x_i)/y_i \quad \hbox{elliptic} \\ 
 \end{array} \right.,\label{ctZibb} \end{equation}
which, in turn, implies (by differentiation) the following constraint between the $\Dim$ and $\Dik$ (hence, between the gradients $m'_{qi}$ and $k'_{qi}$)
\begin{equation} \Dim\left[1-\HH_{qi}c(t_i-\tbb^{(0)})\right] = \Dik\left[1-\frac{3}{2}\HH_{qi}c(t_i-\tbb^{(0)})\right].\label{ctbbDmDk} \end{equation}
As a consequence of (\ref{ctZibb}), a simultaneous big--bang allows us to prescribe only one of the two functions $m_{qi}$ and $k_{qi}$, the other one must be found by solving the algebraic constraint (\ref{ctZibb}) or solving (\ref{ctbbDmDk}) as a differential equation for $\dd k_{qi}/\dd m_{qi}$ (the function $R_i(r)$ remains free and can be fixed as a radial coordinate gauge). Since we need both functions $m_{qi}$ and $k_{qi}$ to solve (\ref{EVql}) or (\ref{EVqlsig}), finding previously a numeric solution of the algebraic constraint (\ref{ctZibb}) is not a problem (it is part of the work needed to provide appropriate initial conditions). However, assuming that $m_{qi}$ is selected and bearing in mind that
\bse\ba  0< Z_h(x_i)\qquad\hbox{for}\quad x_i>0,\label{limZh}\\
0< Z_e(x_i) <\pi \quad\hbox{for}\quad 0<x_i<2,\label{limZe} \ea\ese
it is evident that the hyperbolic branch of (\ref{ctZibb}) does not appear to be restrictive at all to find $|k_{qi}|$. For the elliptic branch, a necessary condition for a solution of (\ref{ctZibb}) that yields $k_{qi}$ is
\begin{equation} 0< c(t_i-\tbb^{(0)}) k_{qi}^{3/2} < \pi m_{qi},\label{tbbcons}  \end{equation}
which, again, does not seem to be restrictive (see \cite{KH4} for comparison). Still, since we cannot solve (\ref{ctZibb}) analytically, nor invert (\ref{ctZbb}) to find $L$, the case of a simultaneous big--bang may be easier to handle analytically with the conventional variables $M,\,E$ and $R$ by means of (\ref{hypZR}) or (\ref{ellZR}), since we only need to fix $\tbb=\tbb^{(0)}$ and then prescribe the functions $M$ and $E$ taking care to fulfill the Hellaby--Lake conditions (\ref{noshx1}) and (\ref{noshx2}). Nevertheless, we can still use $m_{qi}$ and $k_{qi}$  to examine qualitatively these regularity conditions in these models through (\ref{Gsimtbb}).

\begin{itemize}

\item Hyperbolic models

We examine (\ref{Gsimtbb}) in the limits $L\approx 0$ and $L\to\infty$ along constant but arbitrary $r$ (so that the time dependence is concentrated on $L$). Consider the following series expansions of $\HH_q c(t-\tbb^{(0)})$:
\bse\ba \HH_q c(t-\tbb^{(0)}) &\approx& \frac{2}{3}+ \frac{x_i^{1/2}}{15}\,L+O(L^{3/2})\nonumber\\\qquad \hbox{for}\quad L\ll 1,\label{Hctbb1}\\
\HH_q c(t-\tbb^{(0)}) &\approx& 1+\frac{2-\ln(2x_iL)}{x_iL}+O(L^{-2})\nonumber\\ \hbox{for}\quad L\gg 1.\label{Hctbb1} \ea\ese
where we used  (\ref{hypZ1a}), (\ref{mkHL2}) and (\ref{ctZbb}). Taking the leading terms in these expansions, $\Gamma$ in (\ref{Gsimtbb}) becomes 
\ba \Gamma \approx 1+\Dim \quad (L\ll 1),\nonumber\\
 \Gamma \approx 1+\frac{3}{2}\Dik\quad (L\gg 1),\nonumber\ea
leading to the following conditions for the fulfillment of (\ref{noshxG})
\begin{equation} \Dik\geq -\frac{2}{3},\qquad \Dim\geq -1.\label{noshxGhbb}\end{equation}
which are the same as (\ref{noshxGh}) with $\tbb'=0$.

\item Elliptic models

The behavior of $\Gamma$ in (\ref{Gsimtbb}) around $L\approx 0$ for $t\approx \tbb^{(0)}$ follows from an expansion similar to (\ref{Hctbb1}), but using $Z_e$ instead of $Z_h$. This yields also $\HH_q c(t-\tbb^{(0)})\approx 2/3+O(L)$, and lead to $\Gamma\approx 1+\Dim$. Considering now $L\approx 0$ but in the collapsing phase, so that $t\approx \tcoll$, we have $\HH_q c(t-\tbb^{(0)})\to-\infty$. Hence, $\Gamma$ in this limit takes the form 
\begin{equation} \Gamma \approx 3|\HH_q|c(t-\tbb^{(0)})\,\left(\Dim-\frac{3}{2}\Dik\right),\end{equation}
while $\Gamma$ for $t\approx \tmax$ takes the form (\ref{Gmax}). Therefore, the conditions for the fulfillment of (\ref{noshxG}) are simply
\begin{equation} \Dim\geq -1,\qquad \Dim-\frac{3}{2}\Dik\geq 0.\label{noshxGebb}\end{equation}
Notice, by comparing with (\ref{tcollr}), that the second condition above is equivalent to $c\tcoll'\geq 0$, hence (\ref{noshxGebb}) are simply (\ref{noshxGe}) with $\tbb'=0$. It is important to notice that a simultaneous big--bang $\tbb'=0$ does not imply a simultaneous collapsing singularity or maximal expansion ($\tcoll'$ and $\tmax'$ are not zero). 

\end{itemize}    

\subsection{Simultaneous maximal expansion and simultaneous collapsing time.}

These elliptic LTB configurations follow by setting the maximal expansion and collapse times, $\tmax,\,\tcoll$, in (\ref{tmc}) to constants. They have been examined in \cite{KH4}, hence the reader can compare the treatment of these cases in this reference with our treatment in this subsection. 

The condition for $\tmax'=0$ in elliptic models follows directly from (\ref{tmc}) as
\begin{equation} c\tmax= ct_i +\gamma_0,\qquad \pi - Z_e(x_i) = \gamma_0 y_i,\label{simtmax}\end{equation}
where $\gamma_0$ is a positive constant and $x_i,\,y_i$ and $Z_e$ are given by (\ref{ellZ1a}) and (\ref{xye}). As in the case $\tbb'=0$, we can only prescribe one of $m_{qi}$ and $k_{qi}$, with the other one obtained by solving numerically the constraint (\ref{simtmax}). If $m_{qi}$ is prescribed, then a necessary condition for the existence of a solution for $k_{qi}$ follows directly from (\ref{limZe}) and (\ref{simtmax}):
\begin{equation} 0< \gamma_0 k_{qi}^{3/2} < \pi m_{qi},\end{equation}
which looks like (\ref{tbbcons}) and also does not appear to be restrictive at all. Still, it is necessary to work (\ref{simtmax}) numerically or to make further assumptions on $m_{qi}$ or $k_{qi}$ to get more information. The locus of the big--bang and collapse singularities are now given by inserting (\ref{simtmax}) into (\ref{tmc})
\begin{equation} c\tbb = ct_i+\gamma_0-\frac{\pi}{y_i},\qquad  c\tcoll = ct_i+\gamma_0+\frac{\pi}{y_i},\label{tmc_simtmax}\end{equation}
which indicates a time symmetric location of $\tbb$ and $\tcoll$ with respect to $\tmax$. The gradients of $\tbb$ and $\tcoll$ are
\ba  \frac{c\tbb'}{3R'_i/R_i} &=& \frac{\pi}{y_i}\left(\frac{3}{2}\Dik-\Dim\right),\nonumber\\ \frac{c\tcoll'}{3R'_i/R_i} &=& \frac{\pi}{y_i}\left(\Dim-\frac{3}{2}\Dik\right).\label{tmcr22}\ea
where we used (\ref{xye}) and (\ref{Dadef}).

If $\tcoll'=0$, then (\ref{tmc}) implies the following constraint similar to (\ref{simtmax})
\begin{equation} c\tcoll= ct_i +\epsilon_0,\qquad 2\pi - Z_e(x_i) = \epsilon_0 y_i,\label{simtcoll}\end{equation}
where $\epsilon_0$ is a positive constant. Again, we can only prescribe one of $m_{qi}$ and $k_{qi}$, and obtain the other by solving (\ref{simtcoll}). From (\ref{limZe}) we have $\pi<2\pi-Z_e(x_i)<2\pi$, hence, if we prescribe $m_{qi}$ a necessary condition for finding $k_{qi}$ as a solution of (\ref{simtcoll}) is 
\begin{equation} \pi < \epsilon_0 \frac{k_{qi}^{3/2}}{m_{qi}} < 2\pi,\end{equation}
which, again, does not seem to be restrictive, though further information requires either numerical work on (\ref{simtcoll}) or making assumptions on $m_{qi}$ or $k_{qi}$. The big--bang and maximal expansion times and their gradients are
\bse\ba  c\tbb &=& ct_i+\epsilon_0-\frac{2\pi}{y_i},\nonumber\\
 c\tmax &=& ct_i+\epsilon_0-\frac{\pi}{y_i},\label{tmc_simtcoll}\\  \frac{c\tbb'}{3R'_i/R_i}&=&\frac{2\pi}{y_i}\left(\frac{3}{2}\Dik-\Dim\right),\nonumber\\ \frac{c\tmax'}{3R'_i/R_i} &=& \frac{\pi}{y_i}\left(\Dim-\frac{3}{2}\Dik\right),\label{tmcr33}\ea\ese 

By comparing (\ref{tmcr22}) and (\ref{tmcr33}) with (\ref{noshxGe}), then sufficient conditions to fulfill (\ref{noshxG}) in either case $c\tmax'=0$ and $c\tcoll'=0$ are simply (\ref{noshxGebb}). As with the case $c\tbb'=0$, it is not problematic to work out these cases in solving (\ref{EVql}) or (\ref{EVqlsig}) numerically. As in the case $\tbb'=0$, these cases are easier to handle analytically with the conventional variables, since for a given choice of $M$ and $E$, we have $y_i=|E|^{3/2}/M$, and thus $c\tbb$ follows directly from (\ref{tmc_simtmax}) or (\ref{tmc_simtcoll}). Of course, the three free parameters must comply with the Hellaby--Lake conditions (\ref{noshx2}). 

\subsection{Mixed hyperbolic/elliptic configurations.}

LTB models admit more than one kinematic class in their full radial domain. This type of ``mixed'' configurations can be constructed either by smoothly matching LTB regions with one kinematic class to regions of another (as in \cite{ltbstuff}), or simply by choosing the initial value function $k_{qi}$ (which determines the kinematic class) so that it changes sign in its radial domain. 

A particularly interesting mixed configuration is an elliptic region surrounded by a hyperbolic exterior given by the choice
\begin{equation}  k_{qi}\left\{ 
\begin{array}{l}
  >0 \quad \hbox{for}\quad 0\leq r<r_b,\quad\hbox{elliptic region}  \\
  =0 \quad \hbox{for}\quad r=r_b, \qquad\hbox{interface} \\
  <0 \quad \hbox{for}\quad r>r_b, \qquad\hbox{hyperbolic region}\\ 
 \end{array} \right..\label{ranges} \end{equation}  
The fact that $\Dik\to-\infty$ if $k_{qi}\to 0$ (from its definition, see Appendix \ref{divergingDa}), and thus $\Dk\to-\infty$ for all $t$, signals a potential problem in using the quasi--local variables to study an elliptic/hyperbolic model. However, we notice that $\Dk$ does not appear in (\ref{EVql}) nor in (\ref{EVqlsig}). Also, $\Dik$ only appears in the initial conditions (\ref{initconds2}) in the form $k_{qi}\Dik$, which does not diverge as $k_{qi}\to 0$. Hence, this behavior of $\Dik$ has no consequences for the numerical integration of these evolution equations for these configurations.  

For analytic or qualitative work, the limit $k_{qi}\to 0$ must be handled carefully \cite{ltbstuff}. While regularity conditions for these mixed models follow also from (\ref{Ghe}), even if $\Dik\to-\infty$ as $k_{qi}\to 0$, we cannot simply set $k_{qi}=0$ in (\ref{Ghe}). Instead, we examine $\Gamma$ at constant values of $r$ as $r\to r_b$ (or $k_{qi}\to 0$). The following series expansions around $k_{qi}=0$ hold in this limit
\bse\ba \HH_q\,c(t-t_i) &\approx& \frac{2}{3}\left(1-\frac{1}{L^{3/2}}\right)+O(k_{qi}),\\
1-\frac{\HH_q}{\HH_{qi}} &\approx& 1-\frac{1}{L^{3/2}}+O(k_{qi}), \ea\ese
so that $\Gamma$ in (\ref{Ghe}) takes the form
\begin{equation} \Gamma \approx 1+\left(1-\frac{1}{L^{3/2}}\right)\Dim + O(k_{qi}), \label{Gmixed}\end{equation}
which (up to leading order) is the form of $\Gamma$ in (\ref{Gp}). Hence, $\Gamma$ passes smoothly from its elliptic to its hyperbolic form, approaching the form for parabolic models wherever $k_{qi}\approx 0$ for $r\approx r_b$. 

Since the limit $k_{qi}\to 0$ implies $x_i\to 0$ in either the elliptic ($r<r_b$) or hyperbolic side ($r<r_b$), we should also obtain for $L$ the approximated parabolic form (\ref{par2}) for $r\approx r_b$ from the analytic solutions (\ref{phi_L_h}) and (\ref{ellZ2}) in the limit $x_i\to 0$. Expanding $Z_h(u)$ and $Z_e(u)$ around $u=0$ yields 
\bse\ba Z_h(u) \approx \frac{\sqrt{2}}{3}u^{3/2}+\frac{\sqrt{2}}{6}u^{5/2},\\
 Z_e(u) \approx \frac{\sqrt{2}}{3}u^{3/2}-\frac{\sqrt{2}}{6}u^{5/2},\label{Zh0}\ea\ese
Hence, taking only the leading term in $Z_h$ and $Z_e$ and substituting into (\ref{phi_L_h}) and the expanding phase of (\ref{ellZ2}) yields for both elliptic and hyperbolic sides
\begin{equation} L^{3/2}\approx 1+\frac{3}{2}\sqrt{m_{qi}}c(t-t_i) + O(x_i), \label{Lmixed}\end{equation}
which up to the leading term coincides with (\ref{par2}). We did not consider the collapsing phase in (\ref{ellZ2}) because, from (\ref{tmc}), we have $c\tmax\to \infty$ as $k_{qi}\to 0$, hence the worldline marking the interface $r=r_b$ is contained in the expanding phase for all $t$. 

The conditions for the fulfillment of (\ref{noshxG}) can be obtained jointly for the elliptic and hyperbolic regions, bearing in mind that for $r\approx r_b$ we have $\Dik\to-\infty$ but also $\Gamma$ and $L$ take the forms (\ref{Gmixed}) and (\ref{Lmixed}). For both the elliptic and hyperbolic regions we have as $r\to r_b$
\begin{equation} \frac{c\tbb'}{3R'_i/R_i} \approx \frac{\Dim}{3\sqrt{m_{qi}}},\end{equation}
hence $\Dim\leq 0$ (with $\Dim=0$ only at $r=0$) is a sufficient condition for $c\tbb'<0$ in (\ref{noshxGh}) and (\ref{noshxGe}) to hold in the full radial range (note, from (\ref{Dadef}), that this condition is equivalent to $m'_{qi}\leq 0$). Since $c\tcoll(r)\to\infty$ as $r\to r_b$, the collapsing singularity is contained entirely in the elliptic region, hence the existence of the hyperbolic region for $r>r_b$ has no effect on the condition $c\tcoll'\geq 0$ in (\ref{noshxGe}). 

The only remaining condition is $\Dik\geq -2/3$ in the hyperbolic region. Evidently, this condition does not hold in the limit $r\to r_b$ because $\Dik\to-\infty$, but in this limit $\Gamma$ has the form (\ref{Gmixed}), and so (\ref{noshxG}) is not violated. As long as $\Dik$ remains negative we have $k'_{qi}>0$ (from (\ref{Dadef}) with $k_{qi}<0$), and thus $E'>0$ holds, hence condition (\ref{noshx1}) (equivalent to (\ref{noshxG})) also holds. In the elliptic side, we have $k_{qi}\to 0$ with $k'_{qi}<0$, which implies from (\ref{signEr}) that $E'>0$ and $k_i=\RRi/6<0$ near $r_b$. In fact local spatial curvature $k=\RR/6$ is negative for all times near $r=r_b$, even if $k_q> 0$ all the way up to $r=r_b$. This follows directly by applying the integral property (\ref{propq3}) with $r=r_b$ as integration limit and considering that $k'(t,r_b)\leq 0$ and $k_q(t,r_b)=0$ hold for all $t$:
\begin{equation}k(t,r_b) =\frac{1}{R^3(t,r_b)}\int_0^{r_b}{k'(t,x)\,R^3(t,x) \dd x} <0.\end{equation}
This provides an example of how elliptic dynamics does not (necessarily) imply positive local curvature (see Appendix B).

Notice that if instead of a hyperbolic region in (\ref{ranges}), we have a parabolic region ($k_q=0$) for  $r>r_b$, then the results above still apply to the elliptic region in $0\leq r<r_b$. Hence, we would still have $k<0$ near $r=r_b$ in this region, and thus $k<0$ would also hold in the parabolic region. This is an example of how local spatial curvature can be negative in a parabolic region not containing a symmetry.  

\section{Conclusion and further work.} 

We have provided a comprehensive examination of LTB dust models in terms of quasi--local scalars and their fluctuations, which are covariant objects that can be related to covariant scalars in the ``fluid flow'' or ``1+3'' formalism. The motivation, and contents of the article, together with a summary of previous work, concepts and ideas, have been given in detail in the introduction. As we have shown throughout the article,  the initial value parametrization that emerges from these scalars is useful in analytic, qualitative and numerical studies of the models. The ``fluid flow'' evolution equations for these scalars are an appealing and practical alternative to the analytic solutions conventionally used in the literature, as they are fully general and at the same time technically simple: they can be  effectively handled as ordinary differential equations, and thus can be very handy for a numeric treatment of the models. These evolution equations can also be understood in terms of a gauge invariant and covariant (GIC) perturbation formalism, consistent with the covariant fluid flow approach of Ellis {\it {et al}} \cite{ellisbruni89,BDE,1plus3}, as well as the traditional gauge invariant approach \cite{bardeen}. Under this formalism, the dynamics of LTB models can be cast in terms of the dynamics of spherical, non--linear GIC perturbations on a FLRW background.  

Regarding analytic and qualitative work, the quasi--local scalar representation and its associated initial value description provide an appealing theoretical context to understand the conventional parameters of the models, leading to an understanding and appreciation of previous work under a new perspective. This is specially evident in the fact that all scalars can be expressed by simple scaling laws that can be analyzed qualitatively, as well as in the formulation of regularity conditions (the Hellaby--Lake conditions \cite{ltbstuff,suss02,ltbstuff1}) as restrictions on initial value functions. We have also looked at the problem, proposed by Krasi\'nski and Hellaby \cite{KH1,KH3}, of finding the existence of a unique LTB model that is consistent with the ``mapping'' of arbitrary radial profiles of density or velocity at different cosmic times. Also, we have provided a theoretical context for an Omega parameter for LTB models, which has been defined in the literature as an ansatz \cite{LTBfin,num3}. Given the widespread use of LTB dust solutions for dealing with a wide variety of problems and models in General Relativity and Cosmology, the new approach to these solutions presented here has a significant  potential for applications.    
 
In this article we have mostly introduced, discussed and justified the use of a new scalar representation for studying LTB models. However, the formalism that we have presented is potentially useful in future work using these models, hence we are looking at its direct application in separate articles, on the make, dealing with new results on important theoretical and practical issues, such as: (i) the asymptotic properties and boundary conditions of scalars in the radial direction, (ii) the conditions for constructing LTB models with scalars having radial profiles of  ``clumps'' or ``voids'', (iii) the conditions for the existence of a positive ``back--reaction'' in LTB models, which would  allow us to mimmic the effect of an accelerated cosmic expansion in the context of Buchert's scalar averaging \cite{LTBave2,LTBave3}. Together with these articles under elaboration, we are also conducting a comprehensive numeric study of the models, based on the evolution equations derived here. Finally, as shown by \cite{bolejko}, the extension of the formalism of quasi--local scalars can be suitably modified in order to apply it to the quasi--spherical Szekeres spacetimes. The main  justification of the present article is then to motivate and induce researchers to consider the introduced variables in the study and application of LTB models, as well as to serve as the theoretical reference for the use of these variables.

\begin{appendix}
\section{Regularity issues of the new variables. }

The regularity of LTB models in the conventional variables has been extensively discussed in the literature~\cite{kras,kras2,ltbstuff,suss02,ltbstuff1}. We examine this issue in terms of the quasi--local variables and their fluctuations.\\

\subsection{Symmetry centers.}\label{symmetry_centers}

We have only considered in this article LTB models having (at least) one symmetry center, which is a regular timelike  worldline corresponding to a fixed point of the rotation group SO(3). This is a sufficient condition for integrals in (\ref{aveq_def}) to be finite in a domain $\vartheta(r)$ defined by (\ref{etadef}). The symmetry center can be marked as $r=0$ (and $r=r_c$ if there is a second one). The following conditions hold: $R(t,0)=\dot R(t,0)=0$.
Considering (\ref{Ldef}) and (\ref{Gammadef}) and $R'\to 1$ as $r\to 0$, we can mark the symmetry center by a zero of $R_i(r)$ (distinct from the locus of a curvature singularity: $L=0$) \cite{suss02}. Hence
\begin{equation} L(ct,0) >0,\qquad  \Gamma(ct,0) =1.\end{equation}
Since $E(0)=M(0)=M'(0)=E'(0)=0$, then $m_{qi}(0)=m_i(0)$ and $k_{qi}(0)=k_i(0)$ must hold, and thus $\HH_{qi}(0)=\HH_i(0)$. For any scalar $A$ and its dual function $A_q$ we have  
\begin{equation} A(t,0)=A_q(t,0),\qquad A'(t,0)=A_q'(t,0)=0,\end{equation}
The same conditions hold at a second symmetry center $r=r_c$ in closed elliptic models (see Appendix \ref{topology}). Notice that a central singularity is now associated with $L(ct,r)=0$. 



\subsection{The Riemann tensor.}\label{Riemann}

Regardless of which parameters or variables we may use, the regularity at each spacetime point or surface in LTB models can be characterized by the continuity and finiteness of the Riemann tensor ${\cal{R}}_{({\rm{e}})({\rm{f}})({\rm{g}})({\rm{h}})} ={\cal{R}}_{abcd} {\rm{e}}_{({\rm{e}})}^a\,{\rm{e}}_{({\rm{f}})}^b\,{\rm{e}}_{({\rm{g}})}^c\,{\rm{e}}_{({\rm{h}})}^d$, in an orthonormal tetrad basis ${\rm{e}}_{({\rm{e}})}^a$ \cite{kras2,ltbstuff}. Considering the natural tetrad associated with the LTB metric in its form (\ref{LTB2})
\ba  {\rm{e}}_{({\rm{0}})}^a=u^a,\quad {\rm{e}}_{({\rm{1}})}^a=\frac{\sqrt{1-k_{qi} R_i^2}}{\L\,\Gamma\,R'_i}\,\delta^a_r,\nonumber\\ {\rm{e}}_{({\rm{2}})}^a=\frac{1}{L\,R_i}\,\delta^a_\theta,\quad {\rm{e}}_{({\rm{3}})}^a=\frac{1}{L\,R_i\,\sin\theta}\,\delta^a_\phi,\label{tetrad}\ea
the nonzero Riemann tensor basis components are given readily in terms of the new variables $m_q$ and $\Dm$ by
\ba  {\cal{R}}_{(0)(1)(0)(1)}=3m-2m_q=m_q(1+3\Dm),\nonumber\\ {\cal{R}}_{(1)(2)(1)(2)}=3m-m_q=m_q(2+3\Dm),\nonumber\\
 {\cal{R}}_{(0)(2)(0)(2)}=m_q,\quad {\cal{R}}_{(2)(3)(2)(3)}=2m_q.\nonumber\\
\label{Riemann2}\ea 
These basis components also involve other scalars like $\HH,\,\HH_q,\,k,\,k_q$ and $\Dh,\,\Dk$, which are  related to $m,\,m_q$ and $\Dm$ by the constraints (\ref{mkHL1}), (\ref{mkHL2}), (\ref{HdH}) and the Hamiltonian constraint (\ref{cHam_13}) for the local covariant scalars
\begin{equation}\HH^2=2m-k+\Sigma^2=2m-k+(\HH_q\Dh)^2.\label{Hconstr}\end{equation}
Clearly, it is sufficient for the regularity of LTB models based on (\ref{Riemann2}) that all the involved scalars (not only $m$ and $m_q$) are bounded and continuous, but it is not a necessary condition, since $\HH$ and $\HH_q$ could diverge because $k$ and $k_q$ diverge, with $m$ and $m_q$ remaining bounded. Also, we note that the relative fluctuations $\Da$ might diverge in some cases when $A$ and $A_q$ are bounded.

\subsection{Topology of the space slices.}\label{topology}       

Given the existence of (at least) one symmetry center, the admissible topologies (homeomorphic classes) of the $\T[t]$ are

\begin{itemize}

\item {\underline{``Open models''\,:\,\,$\T[t]$ homeomorphic to $\mathbb{R}^3$}}. There is only one symmetry center, at $r=0$. As a consequence of the regularity condition (\ref{noshx}), $R'>0$ and $E>-1$ must hold for all $r$, and so this topology is compatible with regions or models of all kinematic classes (hyperbolic, parabolic or elliptic with $-1<E<0$).  We discuss the asymptotic radial regime associated with this topology in a separate article.   

\item {\underline{``Closed models''\,:\,\,$\T[t]$ homeomorphic to $\mathbb{S}^3$}}.
There are two symmetry centers, at $r=0$ and $r=r_c$. Since $R(t,0)=R(t,r_c)=0$ for all $t$, then there must exist a turning value  of $R$ so that $R'(\rtv)=0$ where $0<\rtv<r_c$.

\end{itemize}

\noindent
Since (\ref{noshx}) is valid in each $\T[t]$ and $\FF=\sqrt{1+E}$ does not depend on $t$, a zero of $\FF$ (characteristic of $\mathbb{S}^3$ topology) or the fulfillment of $\FF>0$ (characteristic of $\mathbb{R}^3$ topology) will be common to all $\T(t)$. Hence, all $\T[t]$ belong to the same topological class and thus
\begin{equation}\hbox{sign}[R'(t_i,r)]=\hbox{sign}[R'(t_j,r)],\label{signRr}\end{equation}
must hold for every $r$ and for any arbitrary pair of distinct hypersurfaces $\T[t_i],\,\T[t_j]$. As consequence, (\ref{noshx}) can be given as an initial condition 
\begin{equation}\hbox{sign}\,(R'_i) = \hbox{sign}\, M' = \hbox{sign}\,\FF,\label{RirF}\end{equation}
specified on an arbitrary fiducial or ``initial'' hypersurface $\T[t_i]$.

\subsection{Possible blowing up of the $\Da$.}\label{divergingDa} 

From their definition (\ref{Dadef}), it is evident that relative fluctuations $\Da$ will diverge under certain regular conditions ({\it{i.e}} continuous and finite $A$ and $A_q$) if in the integration domain $\vartheta(r)$ of (\ref{aveq_def}) $A_q$ has a zero that is not a common same order zero of $A$. This situation occurs in various situations that do not violate regularity, such as the zero of $\HH_q$ at $t=\tmax$ as elliptic models or regions pass from expansion ($\HH_q>0$) to collapse ($\HH_q<0$), or in mixed elliptic/hyperbolic configurations where $k_q$ passes from positive to negative at some comoving radius $r=r_b$ (see section XIV D). However, the quantities $\Dm,\,\Dk$ and $\Dh$ in all basis components in (\ref{Riemann2}) only appear in terms such as $m_q\Dm,\,k_q\Dk$ and $\HH_q\Dh$, which do not diverge at a zero of $m_q$,\, $k_q$ or $\HH_q$. Hence, the blowing up of $\Dh$ and $\Dk$ because of zeroes of $k_q$ or $\HH_q$ does not affect the regularity of the solutions, as conveyed by the continuity and finiteness of (\ref{Riemann2}). Nevertheless, in these cases it will be preferable to use $A_q\Da$, instead of $\Da$,  for studying the time evolution of scalars (which justifies the use of (\ref{EVqlsig}) over (\ref{EVql}) for numeric work involving expanding/collapsing regions or models). 

\subsection{Restrictions of the radial range due to a curvature singularity.}\label{restrictions}

When we introduced the quasi--local scalars in (\ref{aveq_def}), we assumed that the integration range $\vartheta(r)$ defined by (\ref{etadef}) was fully regular. However, in general, the coordinate locus (\ref{Lzero}) of the central singularity (expanding or collapsing) is not simultaneous ({\it i.e.} not marked by $t=\tbb=$ const., so that $\tbb'\ne 0$). Consider the case of an expanding non--simultaneous singularity (big bang), marked by the curve $[\tbb(r), r]$, with $\tbb'\leq 0$, in the $(ct,r)$ coordinate plane, so that $L(\tbb(r), r)=0$ (the collapsing singularity is analogous). The hypersurfaces $\T[t]$ for $t\leq \tbb$ are only regular for the semi open subset
\begin{equation} \bar\vartheta(r)\equiv \{x\,|\, r_{\tiny{\textrm{bb}}} < x \leq r\} \subset \vartheta(r),\label{restr_eta} \end{equation}
where $r_{\tiny{\textrm{bb}}}$ is the intersection of $\tbb(r)$ and the constant $t$ value associated with the $\T[t]$. As a consequence, the integration range in the definition (\ref{aveq_def}) for these $\T[t]$ must be $\bar\vartheta(r)$, not $\vartheta(r)$. For a collapsing singularity in elliptic models we have exactly the same situation, but the involved hypersurfaces are those with $t\geq \tcoll(r)$, with the lower radial bound given by $r=r_{\tiny{\textrm{coll}}}$ marking the intersection of $\tcoll(r)$ and the $\T[t]$.  

However, this range restriction has no consequence in the definition and usage of the quasi--local scalars $A_q$, as the involved integrals can be treated simply as standard improper integrals. We define at each $\T[t]$ with $t\leq \tbb(r)$ the incumbent integrals with their lower integration limit as $y =r_{\rm{bb}}+\epsilon$, for an arbitrarily small $\epsilon>0$, and then obtain the limit as $\epsilon\to 0$. Off course, since scalars like $m,\,k,\,\HH$ diverge in this limit, $m_q,\,k_q$ and $\HH_q$ might diverge as well, but the functions are well defined and behaved in the range $\bar\vartheta(r)$.  This restriction only prevents the $A_q$ from taking values $r<r_{\rm{bb}}$ for $t\leq \tbb(r)$, and so all results that involve these scalars can be trivially extended to include hypersurfaces $\T[t]$ for $t\leq \tbb(r)$ and $t\geq \tcoll(r)$.

\section{Local vs. quasi--local spatial curvature.} 

The kinematic class of LTB models is determined by the sign of the initial quasi--local spatial curvature $k_{qi}$ (or $E=-k_{qi} R_i^2$). From (\ref{mkHL1}), the sign of $k_{qi}$ determines the sign of $k_q$. It is evident that parabolic models (containing a symmetry center) are spatially flat, as $k_{qi}=k_i=0$ trivially implies $k_q=k=0$. However, it is not obvious if a given sign of $k_q$ in regular hyperbolic or elliptic LTB models also determines the sign of the local spatial curvature $k=\RR/6$. In order to examine this point, we rewrite (\ref{slaw2}) as
\begin{equation}k = k_q\,\left[1+\Dk\right] = \frac{k_{qi}}{3\Gamma\,L^2}\left[\Gamma+3\left(\Dik+\frac{2}{3}\right)\right].\label{k1}\end{equation}
We examine the relation between $k_q$ and $k$ and their initial values for hyperbolic and elliptic models.\\ 

\noindent
{\underline{Negative spatial curvature: hyperbolic models}}

\smallskip

In hyperbolic models or regions containing a center, $\Gamma>0$ and $\Dik\geq -2/3$ hold everywhere (conditions (\ref{noshxG}) and (\ref{noshxGh})), therefore, as a consequence of (\ref{mkHL1}) and (\ref{k1}), we have:
\begin{equation} k_q < 0 \quad \Rightarrow \quad k<0,\label{kkqhyp}\end{equation}
Hence, since all regular hyperbolic models or regions comply with $k_q<0$, then all these models or regions have also negative local spatial curvature. As explain further below, the converse is not true, as local spatial curvature can be negative in certain elliptic regions in which $k_q\geq 0$. \\

\noindent
{\underline{Elliptic models and positive spatial curvature.}}

The relation between $k_q$ and $k$ is more complicated in elliptic models or regions, since standard regularity (conditions (\ref{noshxG}) and (\ref{noshxGe})) do not place a lower bound on $\Dik$. Hence, the possibility that $\Dik<-2/3$ occurs cannot be ruled out, and so $k<0$ can happen in regions where $k_q>0$. It is straightforward to show from (\ref{MEDi}) that $\Dik\leq -2/3$ can only occur in a regular elliptic model or region ($\Gamma>0$) if $E'=0$ for some $0<\rtv<r$ in a domain $\vartheta(r)$. As a consequence, we have $\Dik>-2/3$ for all regular elliptic models in which $E'\leq 0$ holds for all $r$ (with $E'=0$ only at the symmetry center), and so (\ref{k1}) implies in this case that $k_q>0\,\Rightarrow\,k>0$ everywhere. 

If there is a zero of $E'$ at some $r=r^*$, then $\Dik< -2/3$ will hold in some regions without violating regularity conditions. Since $-1\leq E\leq 0$ and $E(0)=0$, then for $r\approx 0$ we must have $E'<0$. Thus, the only possible configuration is: $E'\leq 0$ and  $\Dik >-2/3$ for $0\leq r< r^*$, with $E'\geq 0$ and  $\Dik<-2/3$ for $r> r^*$. Further insight into this situation comes from rewriting (\ref{3Ricci1}) as
\begin{equation}  E' =RR'\left(k_q-3k\right)=-4RR'\left[k_q+\frac{k'_q/k_q}{2R'/R}\right].\label{signEr}\end{equation}
If the regularity condition (\ref{RirF}) holds, then $R'/\sqrt{1+E}>0$, and so for  $E'(r^*)=0$ to occur the local curvature $k$ must decrease sufficiently to reach  $k(r^*)=k_q(r^*)/3$, hence $E'>0$ (or $\Dik<-2/3$) leads to further decreasing of $k$, so that $k(r^*)<k_q(r^*)/3$ holds for $r>r^*$. A situation in which $k<0$ occurs with $k_q>0$ can easily be conceived: since $k'$ and $k'_q$ are both monotonously negative, we have a curvature clump and so $k_q>k$, thus, if $k_q$ decays to zero sufficiently fast $k$ might become negative. This happens in the elliptic side of the mixed elliptic/hyperbolic configuration examined in section XIV D.

\end{appendix} 

\vfill
\eject 


\end{document}